\documentclass[twocolumn, twocolappendix]{aastex631}
\usepackage{datatool}
\usepackage{longtable}
\usepackage{amsmath}
\usepackage{float}
\usepackage{afterpage}
\usepackage{booktabs}
\usepackage{wrapfig}
\usepackage{graphicx}
\usepackage[normalem]{ulem}
\usepackage{multirow} 
\usepackage{makecell}

\usepackage{tabularx}

\shorttitle{The SPT-Deep Cluster Catalog}
\begin{document}
\reportnum{DES-2024-0851}
\reportnum{FERMILAB-PUB-24-0919-PPD}

\title{The SPT-Deep Cluster Catalog: Sunyaev-Zel’dovich Selected Clusters from Combined SPT-3G and SPTpol Measurements over 100 Square Degrees}
\author[0000-0002-6633-4519]{K.~Kornoelje$^{1,2,3}$}
 \author[0000-0001-7665-5079]{L.~E.~Bleem$^{3,2,1}$}
 \author{E.~S.~Rykoff$^{4}$}

 \author{T.~M.~C.~Abbott$^{5}$}
 \author{P.~A.~R.~Ade$^{6}$}
 \author{M.~Aguena$^{7}$}
 \author{O.~Alves$^{8}$}
 \author[0000-0002-4435-4623]{A.~J.~Anderson$^{9,2,1}$}
 \author{F.~Andrade-Oliveira$^{8}$}
 \author{B.~Ansarinejad$^{10}$}
% ...

\newcommand{\spitzer}{{\sl Spitzer}}
\newcommand{\newclusters}{10}
\newcommand{\msun}{\ensuremath{M_\odot/h_{70} }}
\newcommand{\fakecib}{dust contamination~}

\newcommand{\minvarnumcount}{500} 
\newcommand{\crossmatchedclusters}{442} 

\newcommand{\minredshift}{0.12}
\newcommand{\medianredshift}{0.74} 

\newcommand{\minmass}{1.0 \times 10^{14}} 
\newcommand{\medianmass}{1.7 \times 10^{14} \msun} 
\newcommand{\maxmass}{8.7} 

\newcommand{\numclustersgteight}{186 } 
\newcommand{\percentclustersgeight}{42}  

\newcommand{\numclustersgtone}{103 } 
\newcommand{\percentclustersgtone}{23}  

\newcommand{\numclustersgtmax}{28 } 
\newcommand{\percentgtmax}{6} 

\newcommand{\cibconatminationhf}{17.4^{+3.1}_{-2.9} \%} 
\newcommand{\cibconatminationlf}{3.7^{+0.7}_{-0.7} \%} 

\newcommand{\confirmrm}{344}
\newcommand{\confirmarchive}{27}
\newcommand{\confirmspitzer}{99}

\newcommand{\completenessmass}{$2.1 \times 10^{14} M_{\odot} / h_{70}$}

\newcommand{\numnewclustersgtfivesig}{102} 
\newcommand{\numnewclustersgtfoursig}{369} 


\newcommand{\bestfittemplowz}{$13.8^{+3.4}_{-2.2}$}
\newcommand{\bestfittempmedz}{$23.6^{+2.1}_{-1.7}$}
\newcommand{\bestfittemphighz}{$21.5^{+1.0}_{-0.9}$}

\newcommand{\highzslpha}{0.18^{+0.20}_{-0.22}}

\newcommand{\ilccilccrossmatchnum}{416 }
\newcommand{\newcilcclustersgtfourhalf}{6 }
\newcommand{\numcilcclusters}{462 }
\newcommand{\numcilcclustersgtfive}{281}
\newcommand{\numcilcclustersconfirmed}{414 }

\begin{abstract}
     We present a catalog of $\minvarnumcount$ galaxy cluster candidates in the SPT-Deep field: a 100 deg$^2$ field that combines data from the SPT-3G and SPTpol surveys to reach noise levels of 3.0, 2.2, and 9.0 $\mu$K-arcmin at 95, 150, and 220~GHz, respectively. This is comparable to noise levels expected for the wide field survey of CMB-S4, a next-generation CMB experiment. Candidates are selected via the thermal Sunyaev-Zel'dovich (SZ) effect with a minimum significance of $\xi = 4.0$, resulting in a catalog of purity $\sim 89 \%$. Optical data from the Dark Energy Survey and infrared data from the \textit{Spitzer} Space Telescope are used to confirm $\crossmatchedclusters$ cluster candidates. The clusters span $\minredshift < z \lesssim 1.8$ and $\minmass  \msun < M_{500c} < \maxmass \times 10^{14} \msun $. The sample's median redshift is $\medianredshift$ and the median mass is $\medianmass $; these are the lowest median mass and highest median redshift of any SZ-selected sample to date. We assess the effect of infrared emission from cluster member galaxies on cluster selection by performing a joint fit to the infrared dust and tSZ signals by combining measurements from SPT and overlapping submillimeter data from \textit{Herschel}/SPIRE. We find that at high redshift ($z>1)$, the tSZ signal is reduced by $\cibconatminationhf$ ($\cibconatminationlf$) at 150~GHz (95~GHz) due to dust contamination. We repeat our cluster finding method on dust-nulled SPT maps and find the resulting catalog is consistent with the nominal SPT-Deep catalog, demonstrating dust contamination does not significantly impact the SPT-Deep selection function; we attribute this lack of bias to the inclusion of the SPT 220~GHz band.
\end{abstract}

\keywords{Large-Scale Structure of the Universe, Galaxy Clusters}

\section{Introduction}
Studies of galaxy clusters can provide valuable insight into fundamental questions in astrophysics and cosmology. In astrophysics, it is well understood that environmental impacts, such as ram pressure stripping and mergers, are fundamental to understanding galactic evolution and the observed differences between galaxy populations in the field and those in clusters \citep[e.g.,][]{Conselice_2014}. In cosmology, galaxy clusters, the most massive collapsed and gravitationally bound structures in the Universe, represent the densest regions in the large-scale matter distribution. As such, the abundances, redshift distribution, and masses of galaxy clusters in our Universe can provide important constraints on cosmological parameters such as the fractional energy density of matter $\Omega_M$, the amplitude of matter density fluctuations $\sigma_8$, the dark energy equation of state parameter $w$, and the sum of the neutrino masses $\sum m_{\nu}$ \citep[][and references therein]{Voit_2005,2011ARA&A..49..409A,Kravtsov_2012}. However, the constraining power of clusters as a cosmological probe is contingent upon: (1) accurate determination of cluster masses, (2) accurate characterization of their selection function, and (3) inclusion of galaxy clusters spanning a wide range of redshifts.

Identification of clusters at millimeter (mm) wavelengths through the thermal Sunyaev-Zel'dovich (tSZ) effect \citep{1972CoASP...4..173S} offers significant advantages for cluster mass estimation 
and the purity of high-redshift cluster samples compared to cluster selection methods in the optical/near-infrared and X-ray \citep[see e.g.,][]{Bleem_2015, 2016, Hilton_2021, Klein_2024}. In the optical and/or near-infrared bands, clusters are identified through overdensities of galaxies \citep[e.g., ][]{1958ApJS....3..211A,2007ApJ...660..239K,2014ApJ...785..104R, Oguri_2017, Gonzalez_2019}, and for low-mass structures hosting few galaxies, projection effects produce a large mass-observable scatter that makes accurate mass estimation difficult \citep[] 
{Costanzi_2018, Grandis_2021,Myles_2021,Wu_2022}. At X-ray wavelengths, clusters are identified by the emission from hot gas ($10^7$ to $10^8$~K) in the intracluster medium (ICM), and this selection is known to be biased toward the detection of cool-core clusters due to a more prominent peak in surface brightness \citep[]{2011A&A...526A..79E, 2017MNRAS.468.1917R,Balzer25}. The reduction of X-ray emission from cosmological dimming also poses a challenge for the detection of high-redshift clusters \citep[]{1999A&A...349..389V, 2004A&A...425..367B, Piffaretti_2011, 2012MNRAS.423.1024M, 2024A&A...685A.106B}. 

In contrast, the signal from the tSZ effect is redshift-independent. It is described by a characteristic spectral distortion of the cosmic microwave background (CMB) due to inverse-Compton scattering of low-energy CMB photons off of high-energy electrons, which takes the form: 

\begin{align}
    \Delta T_{\mathrm{SZ}} &= T_{\mathrm{CMB}} f_{\mathrm{SZ}}(x)\int n_e \frac{k_B T_e}{m_e c^2}\sigma_\mathrm{T} dl \nonumber \\
    &\equiv T_{\mathrm{CMB}} f_{\mathrm{SZ}}(x)y_{\mathrm{SZ}}
\end{align}

\citep{1972CoASP...4..173S}. Here, $n_e$ is the electron number density, $T_e$ is the electron temperature, and $\sigma_{\textrm{T}}$ is the Thomson cross section.  The Compton-$y$ parameter, $y_{\mathrm{SZ}}$, is the total thermal energy of the ICM integrated along the line of sight, and this dependence makes the tSZ signal strength a robust proxy for total cluster mass \citep{2002ARA&A..40..643C,2005ApJ...623L..63M}. The frequency dependence of the tSZ effect, $f_{\mathrm{SZ}}$, is given by:

\begin{equation}
    f_{\mathrm{SZ}}(x) = \left(x \frac{e^x +1}{e^x -1} \right) \left( 1+\delta_{\mathrm{rc}} \right),
\end{equation}
The quantity $x$ depends on frequency, and $\delta_{\mathrm{rc}}$ represents the relativistic correction to the tSZ effect \citep[e.g.,][]{Erler_2018}. At frequencies below the tSZ null, the effect reduces the CMB energy density, providing a distinct signature for identifying galaxy clusters. 

The redshift independence of the tSZ signal also provides a means to detect clusters at high redshifts ($z > 1$), which represents a critical era in astrophysics where significant suppression of star formation occurs in cluster galaxy populations \citep[]{2007A&A...468...33E,2013ApJ...779..138B, Alberts_2016, 2016ApJ...825..113D,2017MNRAS.465L.104N}. The low-redshift cluster environment is dominated by quiescent galaxies whose star formation appears to be quenched by a combination of environmental processes, including mergers and ram-pressure stripping of their hot halos and cool gas disks \citep[]{Bah__2014,2019ApJ...876...40P,kim2023gradualdeclinestarformation}. However, the specific star formation rate of massive galaxies, including the central brightest cluster galaxies, has been shown to increase by $\sim 2$ orders of magnitudes between $z \sim 0$ to $z \sim 1.5$ \citep{2010ApJ...719L.126T,2015A&A...575A..74S, 2017MNRAS.469.1259B}. The fraction of cluster-member galaxies that host an active galactic nucleus (AGN) is also suppressed relative to field galaxies by an order of magnitude at low redshift \citep[]{2010MNRAS.404.1231V,2019MNRAS.484..595M}, but approaches the field value across a similar redshift range ($ z \sim 1.5$) \citep{2006A&A...460L..23P, 2010MNRAS.404.1231V, 2013ApJ...768....1M,2019MNRAS.484..595M}. Such observations suggest that the fraction of cool gas in cluster environments increases with redshift, yet representative samples of galaxy clusters in the high-redshift universe are needed for an accurate, unbiased characterization of this evolution.

A potential challenge for tSZ cluster detection, especially at lower masses, is contamination from correlated astrophysical signals, particularly from sources within the cluster. Two primary contaminants that can hinder the detection of high-redshift clusters at millimeter wavelengths are radio emission from AGN and infrared emission from cluster member galaxies, hereafter referred to as dust contamination. In this analysis, we focus on the impact of the latter, which arises from the fraction of optical and UV stellar emission that is absorbed and re-emitted by dust particles within the interstellar medium. 
The emission of these heated dust particles peaks in the infrared to submillimeter (sub-mm), producing radiation that is well described by a modified blackbody spectrum \citep{Bianchi_2013}. This emission is spatially correlated with clusters and thus partially fills in the tSZ decrement at typical CMB observing frequencies. However, the magnitude of this contamination has yet to be empirically quantified at high redshift \citep[for discussion of low-redshift clusters see e.g.,][]{ Soergel_2017,Erler_2018,2024arXiv240806189Z}. 

Studies of optically selected clusters have suggested that \fakecib could result in a systematic underestimation of tSZ cluster detections \citep{Saintonge_2008,2021MNRAS.502.4026F}, which may produce bias in cosmological constraints  that can be further exacerbated at the highest redshifts due to increased star formation in clusters at $z \sim 2$ \citep{Brodwin_2013}. Previous cosmological studies that used high-redshift, SZ-detected clusters have leveraged dedicated weak lensing observations to place constraints on the evolution of the tSZ observable-mass scaling relation at high redshift, thus marginalizing the impact of the \fakecib on cosmology \citep[]{Schrabback_2021,Zohren_2022}. The sensitivity of upcoming CMB experiments such as CMB-S4 \citep{abazajian2019cmbs4sciencecasereference} will extend the population of SZ-selected clusters to lower masses, making it increasingly important to accurately assess the impact of \fakecib on cluster selection.

Multi-wavelength cluster searches are one effective solution to mitigate the impact of contamination from astrophysical sources on the tSZ signal. At sub-mm wavelengths, emission from \fakecib dominates the tSZ signal by an order of magnitude in flux density \citep[e.g.,][]{Soergel_2017}, making maps of the sub-mm sky a valuable tracer of the spectral energy distribution of dust \citep{2021A&A...653A.135O}. Measurements from mm to sub-mm wavelengths at the locations of galaxy clusters can be used to constrain models of \fakecib and its frequency dependence; this enables the removal of dust from multi-frequency maps, enhancing the detection of tSZ cluster signals that might otherwise be obscured \citep{2024arXiv240806189Z}.

In this work, we use observations of the SPT-Deep field, a 100 deg$^2$ patch of the Southern sky centered at right ascension (R.A.) 0h, declination (Dec.)~$-55^\circ$ from the South Pole Telescope (SPT; \citealt{2011PASP..123..568C})---including data from SPTpol \citep{2012SPIE.8452E..1EA}  and SPT-3G, \citep[]{2014SPIE.9153E..1PB,Sobrin_2022}---and Herschel/SPIRE \citep{2010A&A...518L...3G}. The combined survey data allows us to probe cluster selection at half the limiting mass threshold of previous studies, while the inclusion of 220 GHz and higher-frequency bands allows us to test for dust-induced biases in cluster selection. 

This work is organized as follows: in Section \ref{sec:obs_data_red}, we describe the observations and the processing of the datasets used in cluster finding. In Section \ref{sec:finding_clusters}, we discuss the process of cluster candidate selection. In Section \ref{sec:source_sub}, we describe the handling of point source contamination by radio galaxies in the SPT maps.  In Section \ref{sec:external_datasets_characterization}, we describe the optical and infrared datasets used to confirm and characterize redshifts and galaxy richnesses of cluster candidates. In Section \ref{sec:simulations}, we detail the production of SPT-Deep-like simulated maps used to estimate the purity of the cluster catalog. In Section \ref{sec:min_var_catalog}, we describe the main cluster catalog and compare the results with external cluster catalogs. In Section \ref{sec:CIB}, we investigate the impact of \fakecib on our results and describe the construction of a dust-nulled tSZ cluster catalog. We end with our main conclusions 
in Section \ref{sec:conclusion}.

In this work, where applicable, we assume a fiducial $\Lambda$CDM cosmology with $\sigma_8 = 0.80$, $h = 0.70$, $\Omega_b = 0.046$, $\Omega_m = 0.30$, $n_\mathrm{s}(k_{\mathrm{s}} = 0.002) = 0.972$, and $\sum m_{\nu} = 0.06 e$V. Cluster masses are reported in terms of $M_{500c}$, which is defined as the mass enclosed within a radius, $r_{500c}$, at which the average enclosed density is $500\times$ the critical density at the cluster redshift.

\begin{table}[ht]
    \centering
    \scriptsize 
    \begin{tabular}{p{3.6cm} p{2.5cm} p{1.3cm}}
    \toprule
    \textbf{Survey} & \textbf{Depth} \newline \textbf{(}$\boldsymbol{\mu}$\textbf{K-arcmin)}
\footnote{Unless otherwise specified, temperature units are in ${\rm \mu K_{CMB}}$, referring to equivalent fluctuation in the CMB temperature, i.e., the temperature fluctuation of a 2.73 K blackbody. We omit the CMB subscript for simplicity.} & \textbf{Angular \newline Resolution} \\
    \midrule
    \textbf{SPT-3G (2019--2023)} \\ 
    95~GHz \newline 150~GHz \newline 220~GHz & 3.2 \newline 2.6 \newline 9.0 & 1.6' \newline 1.2' \newline 1.1' \\
    \midrule
    \textbf{SPTpol 500d (2013--2016)} \\ 
    95~GHz \newline 150~GHz & 11.3 \newline 5.2 & 1.7' \newline 1.2' \\
    \midrule
    \textbf{SPTpol 100d (2012--2013)} \\ 
    95~GHz \newline 150~GHz & 13.2 \newline 6.2 & 1.7' \newline 1.2' \\
    \midrule
    & \textbf{Depth} \newline \textbf{(}\textbf{MJy/sr-arcmin)} & \\
    \midrule
    \textbf{\textit{Herschel}/SPIRE (2012)} \\ 
    600~GHz \newline 857~GHz \newline 1200~GHz & 0.055 \newline 0.067 \newline 0.083 & 36.6" \newline 25.2" \newline 18.1" \\
    \bottomrule
    \end{tabular}
    \caption{The noise levels, resolution, and observing years for the four survey fields used to construct the SPT-Deep minimum-variance and/or dust-nulled cluster catalogs: SPT-3G, SPTpol 100d, SPTpol 500d, and \textit{Herschel}/SPIRE. Depths are quoted in $\mu$K-arcmin at $\ell$ = 4000--5000 for the SPT fields and in MJy/sr-arcmin for the \textit{Herschel}/SPIRE maps.}
    \label{table:survey_table}
\end{table}

\section{Observations and Data Reduction}
\label{sec:obs_data_red}
\begin{figure*}
  \centering
  \includegraphics[width=6in]{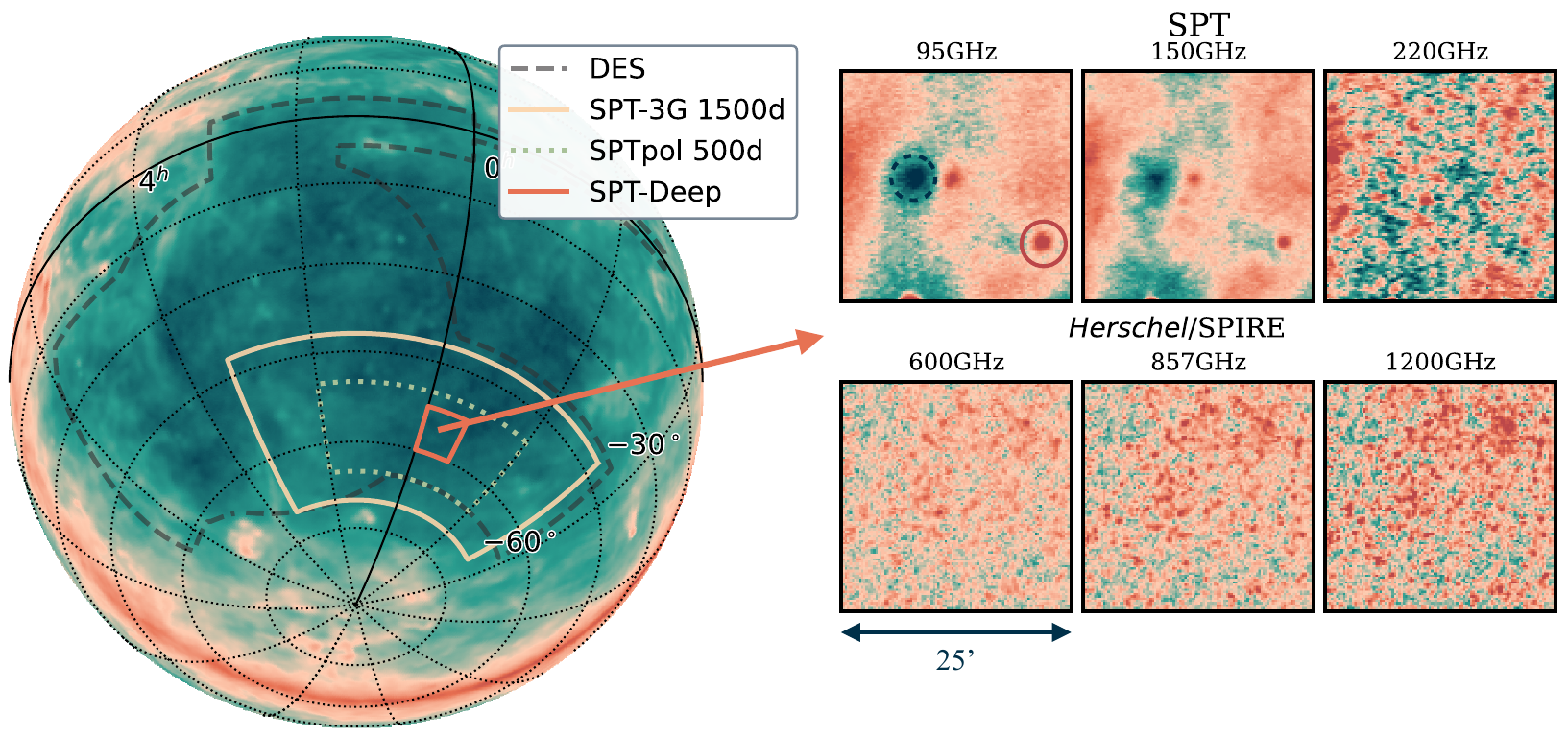}
  \caption{Footprints of the surveys used in the construction of the SPT-Deep cluster catalog. The cluster sample is created using data from SPT-3G (yellow), SPTpol (green), and  \textit{Herschel}/SPIRE (orange, deep field region). Optical/near-infrared imaging from the Dark Energy Survey (gray-dashed) and infrared \textit{Spitzer} data (orange, deep field region) cover the SPT-Deep field and are used to confirm a significant fraction of cluster candidates presented in this work. These survey outlines are overlaid on top of the IRAS 100 $\mu$m dust map from \citet{1998ApJ...500..525S}. On the right, we show 25 arcminute cutouts of the SPT and \textit{Herschel}/SPIRE fields. Top: SPT-3G field at 95, 150, and 220~GHz. Bottom: Cutouts of the \textit{Herschel}/SPIRE fields at 600, 857, and 1200~GHz. In dashed blue, we circle the galaxy cluster SPT-CL J2316-5453, and in solid red, we highlight a source with a flux of $\sim 8$ mJy measured at 95~GHz in the SPT-3G maps.}
  \label{fig:footprint}
\end{figure*}

The SPT-Deep field is a 100 deg$^2$ patch of sky ($10^{\circ} \times 10^{\circ}$) located in the southern hemisphere centered at R.A. 23h 30m and Dec.~$-55^\circ$. This field has been surveyed at multiple wavelengths spanning the radio to X-ray. Millimeter-wave data at 95, 150, and 220~GHz is obtained from two generations of SPT instruments: SPTpol \citep{2012SPIE.8452E..1EA} and SPT-3G \citep[]{2014SPIE.9153E..1PB,Sobrin_2022}. Observations in the sub-mm are acquired from the \textit{Herschel}/SPIRE mission \citep{2010A&A...518L...3G} at wavelengths of 250, 350, and 500 $\mu$m (approximately 1200, 857, and 600~GHz). Cluster confirmation and follow-up is done using a combination of optical data from the Dark Energy Survey (DES) and infrared data from the \textit{Spitzer} Space Telescope \citep[]{2004ApJS..154...10F,Ashby_2014}. The main instruments and data sets used in this analysis are detailed below, with mm- and submm-wave survey specifications detailed in Table \ref{table:survey_table}. The survey footprints and cutouts are shown in Figure \ref{fig:footprint}. External datasets are further characterized in Section \ref{sec:external_datasets_characterization}. 

\subsection{Millimeter SPT Data}
The South Pole Telescope \citep[SPT,][]{2011PASP..123..568C} is a 10-meter-diameter telescope located at the National Science Foundation’s Amundsen-Scott South Pole Station. The instrument's 10-meter aperture enables a diffraction-limited angular resolution of $\sim$1 arcmin at 150~GHz which is well-suited for the angular size of high redshift galaxy clusters.

SPTpol was the second instrument deployed on SPT. It consisted of 1536 detectors: 1176 configured to observe at 150~GHz and 360 configured for 95~GHz. The first SPTpol observing season, and a portion of the second season, were spent observing the SPT-Deep field. Maps made from the weighted sum of all SPTpol observations of this field have a rough noise level of 6.2 $\mu$K-arcmin at 150~GHz and 13.2 $\mu$K-arcmin at 95~GHz, which is roughly a factor of 2-3 lower than that of the previous generation maps from  SPT-SZ \citep{Bleem_2015}. We note that the SPTpol 100d maps used in this analysis are slightly shallower than \citet{Huang_2020} as they were reprocessed to have the same filtering properties as the SPTpol 500d maps. The following four years were dedicated to observing the SPTpol 500d field centered at R.A. 0h and Dec. $-57.5^\circ$ which covers the entire SPT-Deep field. The noise levels of the SPTpol 500d field are comparable to those of the SPTpol 100d field, with a rough noise level of 5.2 $\mu $K-arcmin at 150~GHz and 11.3 $\mu$K-arcmin at 95~GHz. Details of the SPTpol 500d map construction can be found in \citet{Bleem_2024}, hereafter B24. 

The third-generation receiver, SPT-3G, began observations in 2017. In this analysis, we make use of SPT-3G data collected between 2019 and 2023. The SPT-3G main survey covers a 1500 deg$^{2}$ footprint centered at R.A. 0h and Dec. $-56^\circ$, which completely overlaps the SPT-Deep field. With $\sim$16,000 detectors observing at 95, 150, and 220~GHz, the order of magnitude increase in the number of bolometers enables a proportional reduction in noise \citep{Sobrin_2022}. Combined, the coadded SPTpol 100d, SPTpol 500d, and SPT-3G maps have rough noise levels of 3.0, 2.2, and 9.0 $\mu$K-arcmin at 95, 150, and 220~GHz respectively. These noise levels are comparable to those predicted for the CMB-S4 wide survey \citep{abazajian2019cmbs4sciencecasereference}.

\subsection{SPT Map Making Process}
\label{spt_map_making}
The process for converting time-ordered detector (TOD) data into CMB maps in this analysis is optimized for cluster detection and follows established methods from previous studies (see e.g., \citealt{Dutcher_2021} for a more detailed discussion). We provide a brief overview of the SPT-3G map-making process below and refer the reader to \citet{Huang_2020} for the construction of the SPTpol 100d maps, B24 for the construction of the SPTpol 500d maps, and \citet{2013ApJ...771L..16H} for the construction of the \textit{Herschel}/SPIRE maps. 

The SPT scan strategy involves moving the telescope back and forth across the field in azimuth, incrementing in elevation, and repeating the process until the entire field is covered. One scan, or a complete sweep across the SPT-3G field, takes $\sim 100$ seconds to complete, and a full subfield observation is completed in $\sim 2.5$ hours. A series of linear processing steps are performed on TOD data per scan to mitigate and flatten noise in the SPT signal range. These filters include an anti-aliasing Fourier-space filter of the form $e^{(-\ell_x/\ell_0)^6}$  with low-pass cutoff $\ell_0 = 20,000$; a high-pass filter that projects out Fourier modes below $\ell_x = 500$ to remove noise such as atmospheric contamination; and a common-mode filter which removes most of the temperature signal on scales $\ell < \sim 500$. Sources with fluxes exceeding 50 mJy measured at 150~GHz are interpolated over in the TOD. We note that this flux threshold differs from that used in the construction of the SPTpol maps, where sources brighter than 6 mJy at 150~GHz are masked. 

A series of calibration observations are done on the Galactic HII region, RCW38, to relate input detector power to fluctuations in CMB temperature. To prevent degradation in the quality of the final map, several quality assurance checks are done on individual detectors, scans, and observations. Examples of reasons why data may be cut from map construction include irregular TOD data features, errors in telescope pointing information, or errors in data acquisition. An inverse-variance weight is applied to each detector based on its TOD noise, and detectors with weights outside $3\sigma$ of the mean weight are additionally dropped from map construction. A final absolute temperature calibration is applied to the SPT-3G maps which is computed from cross-correlating SPT-3G maps with maps from the \textit{Planck} satellite. 

\subsection{\textit{Herschel}/SPIRE}
The \textit{Herschel Space Observatory} \citep{2010A&A...518L...1P} was a space-based telescope that operated in the far-infrared and sub-millimeter wavelengths. We use data from the \textit{Herschel} Spectral and Photometric Imaging Receiver \citep[SPIRE;][]{2010A&A...518L...3G}, which contained detectors that operated at 600, 857, and 1200~GHz, with approximate beam sizes of 36.6", 25.2", and 18.1", respectively \citep{Viero_2019}. The maps used in this analysis cover a $\sim$90 deg$^2$ patch of the sky centered at R.A. 23h 30m and Dec. $-55^\circ$ and were obtained under an Open Time program (PI: Carlstrom). These maps have noise levels of  0.055, 0.067, and 0.083 MJy/sr-arcmin, at 600, 857, and 1200~GHz, respectively. The observations were conducted using SPIRE fast-scan mode with a scanning speed of 60 arcsec $\mathrm{s}^{-1}$ and are composed of two sets of perpendicular scans. Maps were made from these observations with SMAP \citep[]{2010MNRAS.409...83L,Viero_2013}, an iterative mapmaker optimized to separate large-scale noise from signal. A detailed description of the SPIRE maps can be found in \citet{2013ApJ...771L..16H} and the map-making procedure in \citet{2010MNRAS.409...83L}. Certain pixels in the 1200 and 600 GHz maps were masked due to known instrumental effects resulting in large outlier pixel values; these constitute a negligible fraction of the total map pixels. 

\begin{figure}  
    \centering
    \includegraphics[width=0.47\textwidth]{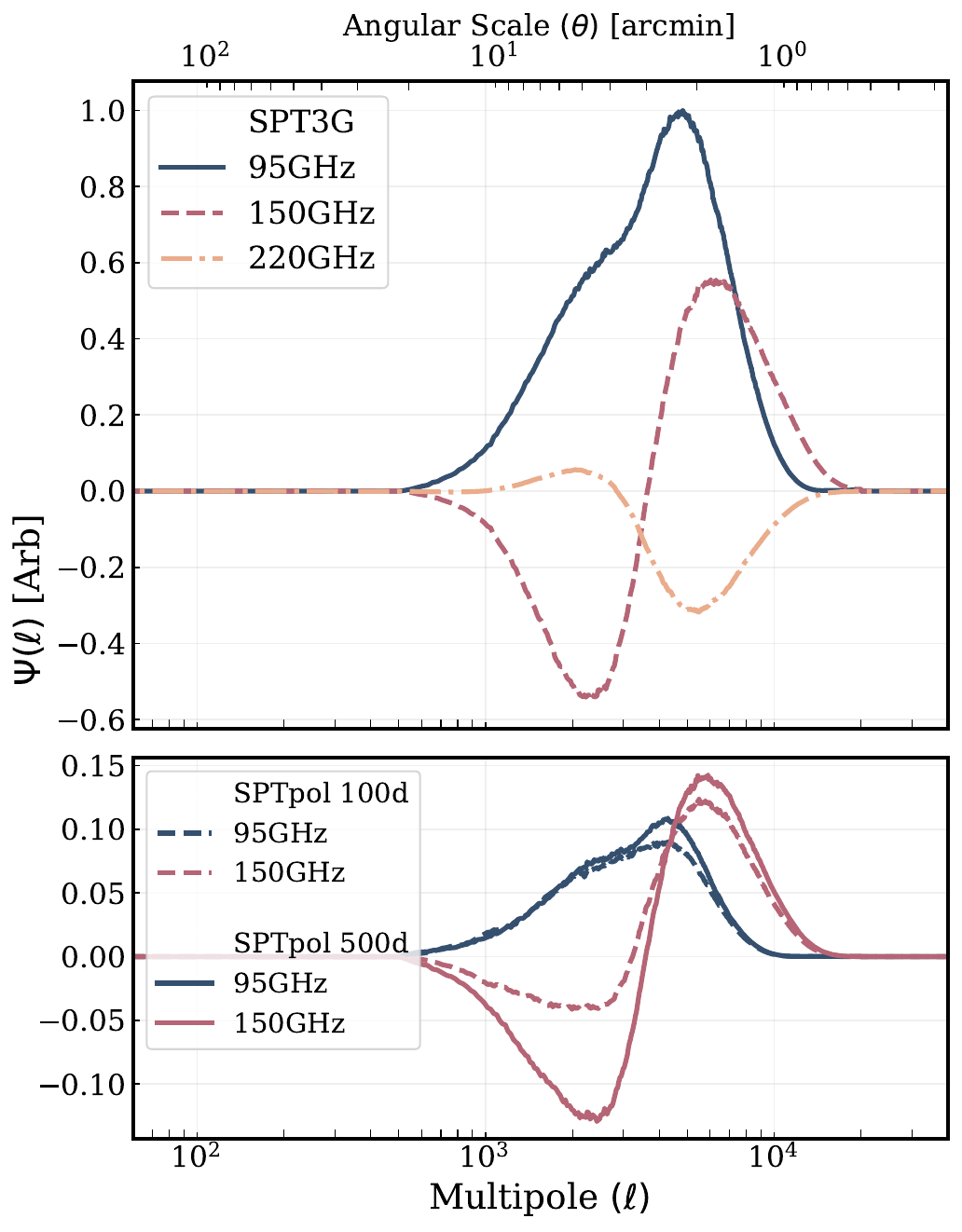} 
    \caption{The azimuthally averaged spatial-spectral matched filter weights for the 0.25' core size beta profile, normalized to the peak of the SPT-3G 95~GHz filter. These weights demonstrate that the lowest-noise SPT-3G CMB maps carry the most weight in cluster detection, followed by SPTpol 100d and SPTpol 500d which carry similar weights.
    \label{fig:psi_figured}}
\end{figure}

\section{Cluster Candidate Identification}
\label{sec:finding_clusters}
In this section, we summarize the process of extracting the tSZ cluster signal from CMB temperature maps. The procedure is similar to those used in previous SPT publications and a more detailed description can be found in e.g.,  \citet{2011ApJ...738..139W} and \citet{2013ApJ...763..127R}. Notably, this analysis constitutes the first SPT cluster catalog produced that includes the 220~GHz frequency band.

\subsection{Sky Model and Minimum-Variance Matched Filter}
At 95, 150, and 220~GHz, the SPT fields are composed of signals stemming from a range of astrophysical sources, each with its own spatial and spectral dependencies. Similar to the techniques described in \citet{2011MNRAS.410.2481R} and \citet{2006A&A...459..341M}, we construct a set of optimal map weights to produce a minimum-variance, unbiased map of the tSZ signal optimized on galaxy cluster scales of a few arcminutes. We assume that each map can be represented as a linear combination of signal and noise components, modeled as:

\begin{multline}
    T(\vec{\theta}, \nu_i) = B(\vec{\theta},\nu_i) * [\Delta T_{\textrm{SZ}}(\vec{\theta}, \nu_i) + N_{\textrm{astro}}(\vec{\theta}, \nu_i)] + \\ N_{\textrm{noise}} (\vec{\theta}, \nu_i).
\end{multline}

The cluster signal is modeled as temperature ($T$) variations as a function of spatial scale ($\vec{\theta}$) and frequency ($\nu$) over the SPT bands relative to the CMB as $\Delta T_{\mathrm{SZ}}$, which is contaminated by emission from astrophysical noise terms, $N_{\textrm{astro}}$. The sources of noise included in $N_{\textrm{astro}}$ are the tSZ and kinetic SZ signal, the Poisson and clustered emission from dusty galaxies with amplitudes obtained from \citet{Reichardt_2021}, and the primary CMB as described by the best-fit lensed Planck 2018 $\Lambda$CDM primary CMB spectrum \citep{Planck2020}. A detailed discussion on astrophysical noise can be found in Section \ref{sec:simulations}. $B(\vec{\theta},\nu_i)$ captures the effect of the beam and map filtering discussed in Section \ref{spt_map_making}. The instrumental noise and residual atmospheric noise are denoted as $N_{\textrm{noise}}$. 

The optimal filter then takes the following form in the Fourier domain:

\begin{equation}
    \psi(l,\nu_i) = \sigma_{\psi}^2 \sum_{j} \textrm{N}_{ij}^{-1}(l) f_{\textrm{SZ}}(\nu_j)S_{\textrm{filt}}(l,\nu_j).
\end{equation}

Here, $\textrm{N}_{ij}^{-1}(l)$ is the inverse astrophysical and instrumental noise covariance matrix which runs over bandpasses $\textrm{(i,\;j)}$ and $S_{\textrm{filt}}(l,\nu_j)$ is the spatial model for the cluster tSZ signal convolved with $B(\vec{\theta},\nu_i)$. The cluster tSZ signal is modeled as an isothermal projected $\beta$-model \citep{1976A&A....49..137C}, with $\beta$ fixed at 1 for the tSZ surface brightness template:

\begin{equation}
    S = \Delta T_0 (1+ \theta ^2 / \theta_c^2)^{-\frac{3}{2}\beta + \frac{1}{2}}.
\end{equation}

The normalization ($\Delta T_{0}$) is a free parameter and we search over a range of core radii ($\theta_c$) from $0.25'$ to $3.0'$ in $0.25'$ steps. As explored in \citet{2010ApJ...722.1180V}, this simplified model of the cluster's gas profile is adequate relative to the resolution of SPT maps. We verify these results in the new, low-level noise regime of the SPT-Deep maps by substituting the cluster pressure profile defined in \citet{Arnaud_2010} in place of our nominal $\beta$-model. This substitution yields consistent results in both the average signal-to-noise of cluster candidates and the number of detected clusters between the two assumed profiles. The predicted variance, $\sigma_{\psi}^2$, of the filtered map, is given by:

\begin{equation}
    \resizebox{0.47\textwidth}{!}{$
    \sigma_\psi^{-2} = \int d^2l \sum_{i,j}f_{\textrm{SZ}}(\nu_i)S_{\textrm{filt}}(l,\nu_i)\textrm{N}_{ij}^{-1}(l)f_{\textrm{SZ}}(\nu_j)S_{\textrm{filt}}(l,v_j).
    $}
\end{equation}
As an example, the resulting frequency-dependent matched filters for the 0.25\arcmin \ $\beta$-model are shown in Figure \ref{fig:psi_figured}. 

We estimate the noise in the resulting filtered map by fitting a Gaussian to the distribution of all unmasked pixel values within $5\sigma$ of the mean in $1.5^{\circ}$ strips to capture any declination-dependent noise variations in the map. The significance of a cluster candidate, $\xi$, is defined as the maximum signal-to-noise ratio measured across all filter scales, and we set the minimum significance threshold for detection to $\xi = 4$ to maintain the purity of the resulting catalog, as spurious detections increase quickly with decreasing $\xi$ thresholds. To control for point-source-related effects that impact the purity of the resulting catalog, we follow previous SPT cluster analyses and: 1) mask (set to zero) a $4{^\prime}$-radius region around every source above a certain flux threshold; 2) exclude all cluster candidates found within an $8^{^\prime}$ radius of these sources and high-significance clusters ($\xi > 6$ measured in SPT-SZ). Following B24, we use a source flux threshold of $6$~mJy at 150~GHz (measured in SPT-SZ data). Again following B24, we mitigate contamination from lower-flux sources using a template-subtraction method discussed in the next section.

\begin{figure}
\includegraphics[width=0.45\textwidth]{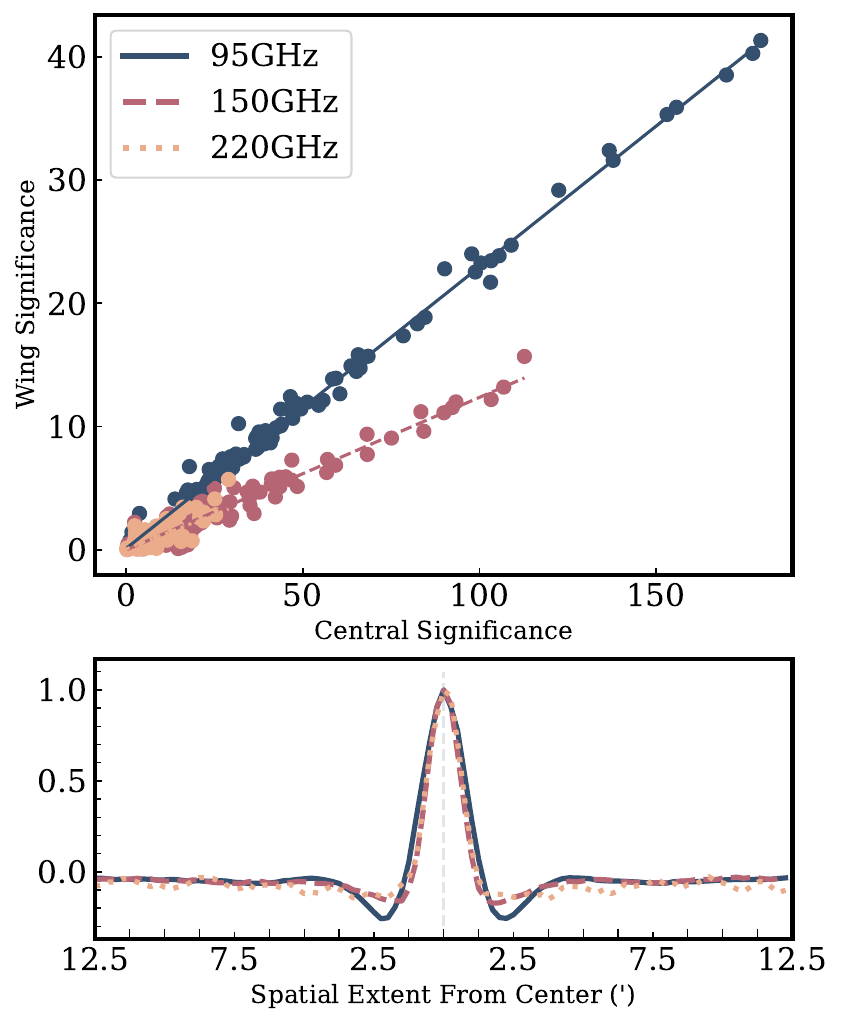}
  \caption{Top: The central significance of point sources plotted against the significance of their deepest ringing wings. The significance is calculated from individual, optimally filtered frequency maps with an additional 0.25' arcminute core size spatial profile applied. Bottom:
  Median stacks of approximately 100 point sources in the SPT-Deep field with fluxes greater than 4~mJy at 150~GHz in the SPT-3G 95, 150, and 220~GHz bands, normalized to the source's peak significance. We find the deepest ringing wings of point sources occur roughly 2 arcminutes away from the central point source location and have roughly $24 \%$ of the source's central significance at 95~GHz. At 150 and 220~GHz, we find this number reduces to $12\%$ and $10\%$, respectively. The significant decrements to the sides of the peak source emission pictured are the source of spurious cluster candidates.}
  \label{fig:source_plot}
\end{figure}

\begin{figure*}
  \centering
  \includegraphics[width=\textwidth]{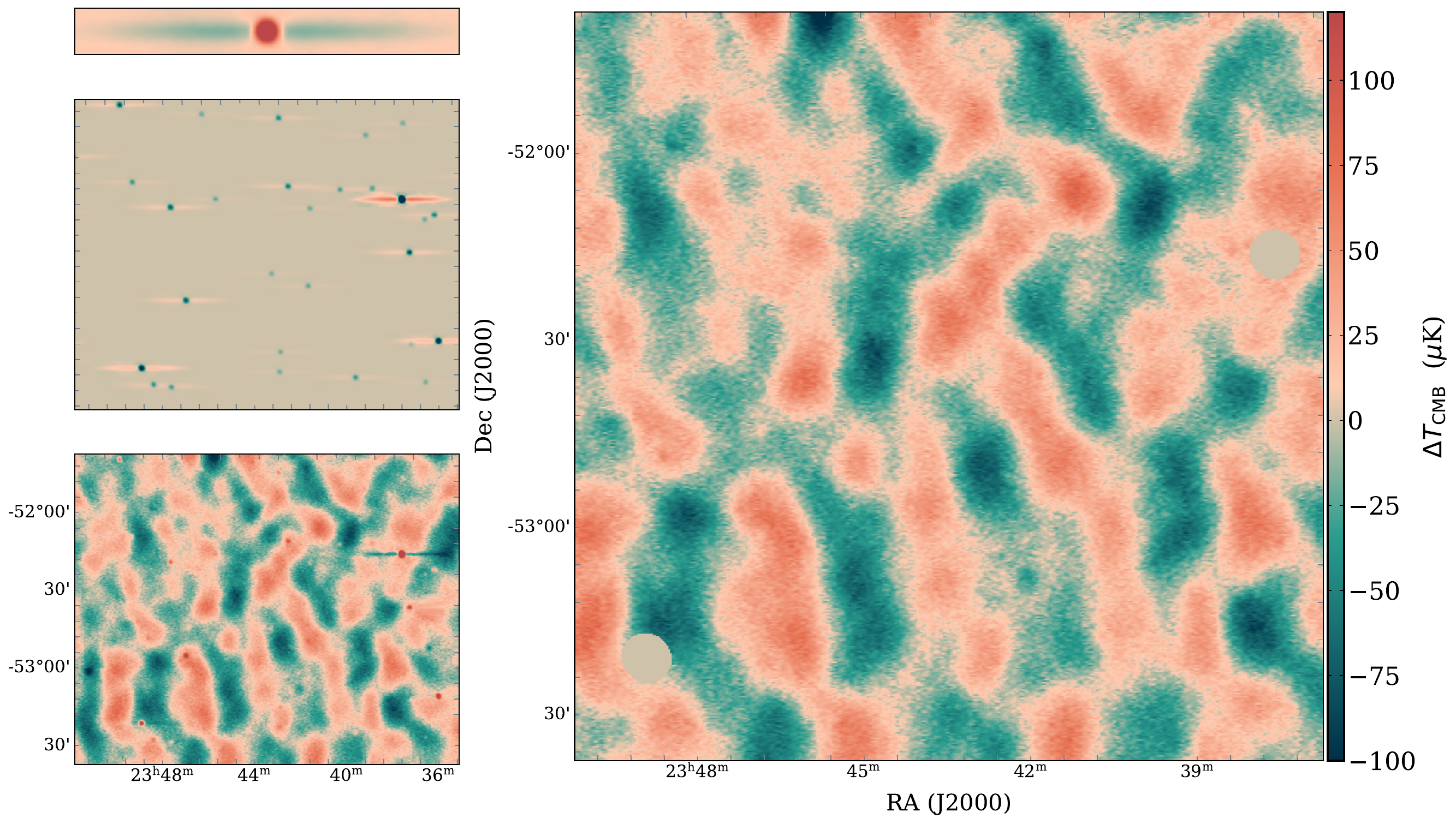}
  \caption{The results of source subtraction as demonstrated on a 2$^\circ$ $\times$ 2$^\circ$ patch ($\sim 4 \%$ of the total map area) of the SPT-3G 95~GHz field. On the left, from top to bottom, we show the template point source, the scaled source subtraction map, and the unfiltered 95~GHz CMB map. On the right, we show the final cleaned map composed of the source subtracted map and a 4'  point source mask for sources brighter than 6~mJy measured at 150~GHz. This masking threshold matches that used in the creation of the SPTpol maps.}
  \label{fig:source_subtraction_map}
\end{figure*}

\section{Source Subtraction}
\label{sec:source_sub}
One challenge to producing a clean catalog of galaxy cluster candidates at mm wavelengths is the contamination from point sources such as bright radio galaxies. The high-pass filter applied to the SPT TOD in mapmaking introduces artifacts around bright point sources in the resulting maps of the sky. The most serious of these for cluster analyses are negative ``wings" along the scan direction, which can lead to false positives in cluster detection (see Figure \ref{fig:source_subtraction_map}). This is the motivation for interpolating over bright sources before TOD filtering, or masking (de-weighting) them in the construction of the filter (see Section~\ref{spt_map_making}).

For many earlier SPT cluster analyses, with higher noise levels, masking all sources above 6 mJy in filtering (as was done with the maps in this work) was sufficient to avoid significant numbers of false positives from point source wings, and the small number of spurious detections were removed by visual inspection. However, the SPTpol data used in \citet{Huang_2020} achieved sufficiently low noise levels that spurious detections from sources below the mask threshold used in mapmaking were identified as a potentially significant source of contamination in future samples. In B24, point source handling was enhanced through the construction of a source template that was used to subtract off emission at previously identified source locations that were not masked in TOD filtering. This effectively eliminated spurious contamination above the selection threshold of $\xi>4$. The minimum significance for source subtraction in B24 was set to sources with a signal-to-noise ratio (SNR) $> 6$ measured at 150~GHz found in a dedicated point source run in the B24 maps. This threshold was found to be sufficient following visual inspection of candidate lists, but no detailed studies were conducted to ensure this was the optimal significance threshold to minimize cluster contamination. As the SPT-Deep field is roughly 2$\times$ deeper than the data used in B24, it is even more important to robustly understand the impact of point source contamination on cluster finding to ensure minimal contamination in the resulting SPT-Deep cluster catalog.

We do this by first empirically quantifying the mapping between the brightness of an emissive source and the spurious decrement created when it is high-pass filtered. We begin by taking cutouts from single-frequency SPT-3G maps that have been spatially filtered with a kernel matched to an 0.25' $\beta$ profile at the locations of previously identified emissive sources. We convert these cutouts into ``units" of significance or SNR by dividing each cutout by the rms in the full filtered map at a similar declination to the cutout (see Section \ref{sec:finding_clusters} for details). This allows us to predict the maximum spurious decrement caused by the combination of the TOD filtering and the cluster-matched filter as a function of source brightness. In the bottom panel of Figure \ref{fig:source_plot}, we show a slice along the scan direction of a median stack of roughly 100 sources with fluxes between 4~mJy and 50~mJy measured at 150~GHz in the SPT-Deep footprint at 95, 150, and 220~GHz. The significant decrements to the sides of the peak source emission are the source of spurious cluster detections. The top panel of Figure \ref{fig:source_plot} quantifies the relationship between the significance of the central (positive) peak and the first (negative) wing.

We find that the significance of wing decrements relative to central source significance is highest at 95~GHz, with a slope of roughly 0.24, while the slope at 150 and 220~GHz is noticeably lower. We therefore set the lower threshold for source subtraction to sources with SNR $ > 5$ measured in a dedicated point-source analysis at 95~GHz. This was chosen to minimize the significance of the source's wings to SNR $ \lesssim 1 $ at 95~GHz while also maintaining high purity in the subtracted source list. This significance threshold corresponds to a flux of roughly $\sim2$~mJy at 95~GHz in SPT-3G, resulting in approximately 750 cleaned sources.

The process of point source cleaning follows techniques in B24 with a few additional updates. First, for the SPTpol field, we construct a point source template by extracting a $6^{\prime} \times 45^{\prime}$ cutout (aligned with the telescope's scan direction and matched to the spatial extent of the brightest unmasked sources) for all sources in the SPT-Deep field with SNR $ > 7$ measured at each map's respective frequency before median-stacking the images. For SPT-3G, because we require a higher-fidelity template (owing to the higher threshold for source masking/interpolation), we follow the same template construction but utilize point source cutouts from the full SPT-3G 1500 deg$^2$ field to reduce the noise in the resulting template. The point source fluxes used to scale the source template are measured in a minimum-variance matched filtered map (discussed in depth in Section \ref{sec:finding_clusters}) with a $\delta$-function (point source) spatial filter. While this controls for differences in source flux between map runs, these measured fluxes can be inaccurate in situations where low-flux sources lie in the ringing wings of higher-flux sources. To circumvent this, we subtract point sources iteratively. High flux sources are subtracted first to eliminate the most problematic ringing wings, and this new, cleaned map is passed through the matched filter again to remeasure source fluxes. This process of subtraction is then repeated for lower flux thresholds until the ringing wings and central source significances lie well below the cluster detection threshold. The various steps of this process are illustrated on a small patch of SPT-3G 95~GHz data in Figure \ref{fig:source_subtraction_map}. 

The treatment of sources in the SPT-Deep map can be summarized as the following: First, sources with SNR $> 5$ at 95~GHz (measured in a dedicated point source run on the SPT-3G maps) but not interpolated over or masked in mapmaking are subtracted (from both the SPTpol and SPT-3G maps) using the template procedure described in this section. Then, a region of radius $4{^\prime}$ is set to zero around sources with fluxes $> 6$ mJy at 150~GHz (measured in SPT-SZ data), and any cluster detections within $8{^\prime}$ of such sources are discarded.

\section{External Datasets and Candidate Characterization}
\label{sec:external_datasets_characterization}
We confirm and characterize cluster candidates by identifying an excess of red-sequence or near-infrared-selected massive galaxies at the location of tSZ cluster candidates. For the majority of cluster candidates at low redshifts ($z \lesssim 1.1$), we use the red-sequence Matched-Filter Probabilistic Percolation (redMaPPer) algorithm \citep{Rykoff_2014,rykoff16}, which relies on optically selected galaxy cluster samples obtained from DES data. However, for high-redshift clusters ($z > \sim1.1$), where the depth of optical data is no longer sufficient to reliably confirm cluster candidates, we employ \spitzer/IRAC data and a modified version of the ``1.6 $\mu$m Stellar Bump Method" (see Section \ref{stellarbump}) for confirmation. Finally, for regions without \spitzer \ coverage or that are partially masked ($>$20\% within a radius of 0.5 Mpc/h  at $z=0.5$) in the DES analysis, we also make use of the DES+WISE optical-infrared cluster catalog of \citet{wen24}. 
The datasets used in this analysis, as well as the specific methods used for cluster identification and redshift estimation, are described in more detail in the subsequent subsections and cited references.

\subsection{External Datasets}
The Dark Energy Survey (DES) is an optical-to-near-infrared imaging survey covering ${\sim5000}$~deg$^2$ of the southern sky with the DECam imager \citep{2015AJ....150..150F} on the 4m~Blanco Telescope at Cerro Tololo Inter-American Observatory. In this work, we make use of DES data in the $griz$ bands acquired from DES-Y6 data, which reach median $10\sigma$ coadded magnitude depths of 24.7, 24.4, 23.8, 23.1,  respectively, in 1.95$"$ apertures \citep{Abbott_2021}. 

We also make use of \spitzer /IRAC imaging \citep{2004ApJS..154...10F} of the SPT-Deep field \citep[SSDF;][]{2013ApJS..209...22A} at 3.6 and 4.5 $\mu$m (henceforth the IRAC [I1] and [I2] bands, respectively) that covers a total area of 94 deg$^2$. The near-infrared imaging with \spitzer \ is sufficiently deep to identify galaxy clusters up to $z \sim 2$. 

\subsection{Cluster Candidate Confirmation, Redshift, and Mass Estimation}
In this section, we provide a brief overview of the methods (redMaPPer, ``1.6 $\mu$m Stellar Bump", archival catalogs) used to confirm cluster candidates and to obtain mass and redshift estimations.

\subsubsection{redMaPPer}
The red-sequence Matched-Filter Probabilistic Percolation (redMaPPer) algorithm \citep{Rykoff_2014,rykoff16} utilizes a maximum-likelihood approach to estimate cluster redshifts based on the cluster galaxy red sequence \citep{2000AJ....120.2148G} alongside galaxy clustering information. To identify optical counterparts for SZ-selected galaxy clusters, redMaPPer is run in `scanning mode' using the tSZ cluster positions as priors. The likelihood of a cluster at each position is then evaluated by measuring the cluster richness $\lambda$---defined as the excess weighted sum of red sequence galaxies within a specified radius, relative to the field galaxy density---as a function of redshift. We adopt the peak richness along the line of sight as our most probable cluster counterpart. 
In constructing the $\lambda$ estimate the optically selected galaxies are filtered and weighted by various filters and corrected for masking and completeness effects. These filters include, among others, a radial filter that adopts a projected Navarro–Frenk–White (NFW) profile \citep{1996ApJ...462..563N} and a color-magnitude filter that weights optical counterparts based upon their consistency with the color-magnitude relation of red-sequence cluster galaxies \citep{2000AJ....120.2148G}. The median redshift uncertainty for redMaPPer confirmed clusters is $\sigma_z/(1+z)$ = 0.006. For more details on the redMaPPer algorithm see \citet{rykoff16}.

\subsubsection{The 1.6 $\mu$m Stellar Bump Method}
\label{stellarbump}
The spectral energy distribution of galaxies older than 10 Myr peaks at a rest-frame wavelength of approximately $1.6 \, \mu \textrm{m}$, where the opacity of the $\textrm{H}^{-}$ ion is minimized (e.g., \citet{1988A&A...193..189J}). This feature provides a distinct marker for photometric redshift estimates of galaxy clusters and contributes to the relative uniformity of the IRAC [I1] - [I2] color in composite stellar populations. The color exhibits a monotonic relationship with redshift across our range of interest ($\sim0.7 < z < \sim1.7$), enabling a reliable mapping between color and redshift \citep[]{1988A&A...193..189J, Sawicki_2002, 2013ApJ...767...39M, 2008ApJ...676..206P, Muzzin_2013, Gonzalez_2019}. As demonstrated in \citet{2010ApJ...721.1056S}, this feature is robust and insensitive to specific choices of stellar population model parameters. We model this color-redshift relation with the GALAXEV package \citep{2003MNRAS.344.1000B} assuming a simple stellar population with a range of metallicities formed in a single burst of star formation at $z = 3$ initialized with a Salpeter initial mass function \citep{1955ApJ...121..161S} which passively evolves following the MILES \citep{2010MNRAS.404.1639V} evolutionary tracks thereafter. 

Following the analysis done in B24, sources selected from the \spitzer \ data are cross-matched with a 1\arcsec \ radius to optically identified counterparts from DES. For accurate redshift estimation using the stellar bump method, low-redshift interloper galaxies must be excluded. These low-redshift galaxies are filtered from the analysis through the application of the $z - \mathrm{[I1]} < 1.6$ cut introduced in \citet{Muzzin_2013}. Following this cut, the single-color $\lambda$-richness estimator introduced in \citet{2012ApJ...746..178R} but tailored to the 1.6 $\mu$m bump features is run in place of the nominal red-sequence model. Richnesses are then computed for each galaxy cluster candidate at redshifts between $z = 0.8$ and $z = 2$, with the cluster's redshift assigned to the value of $z$ that maximizes the cluster richness value $\lambda$. Following \citet{Bleem_2015} and B24, we quote a redshift uncertainty of $\sigma_z/(1+z)$ = 0.035 based on tests with a modest (16 clusters) spectroscopic training set. 

We caution that the highest redshifts are uncertain for several reasons. First, the redshift constraints obtained from the [I1]-[I2] relation degrade substantially outside the range in which this relation is monotonic. 
We thus report $z=1.6$ as a lower redshift limit for galaxy cluster candidates with redshifts estimated at $z=1.6$ or higher.  
Second, while the SSDF catalog includes detections of high-$z$ cluster galaxies, it is a comparatively shallow \textit{Spitzer} survey \citep[5$\sigma$ sensitivity limits at 19 Vega in {[I1]};][]{2013ApJS..209...22A}.  
High-redshift cluster members are faint and the photometric uncertainty can be significant. 
Based on inspection of the SSDF imaging around SPT cluster candidates, we find we need to retain galaxies with color errors as large as $\sigma_\textrm{[I1]-[I2]} =0.4$ to detect the galaxies in the environs of our highest-redshift clusters.\footnote{For reference, the difference in [I1]-[I2] color between $z$=1.2 to $z=$1.6 is 0.3 in our model (see e.g., Fig. 3 in B24). As the redshift estimate is derived from essentially the mean of the color over $\gtrsim$10 galaxies, $\sigma_z$ for the faintest systems is increased by $0.1$  by photometric errors alone.} 
Finally, the richness of the typical SPT-Deep cluster is significantly poorer than our limited spectroscopic calibration set \citep[obtained for SPT-SZ clusters at much higher masses in typically deeper \textit{Spitzer} data, see e.g.,][]{khullar19}. Work is ongoing to significantly increase the spectroscopic calibration sample at $z>1.3$ using new observations from the Magellan/LDSS-3C camera. Future data from targeted follow-up and surveys such as the Vera Rubin Observatory's Legacy Survey of Space and Time \citep[LSST;][]{2019ApJ...873..111I} and Euclid \citep{2024arXiv240513491E} will also improve the photometric measurements of the cluster galaxies.

\subsubsection{Other Redshifts}
The DES analysis includes stringent masking of regions that may have biased photometry (owing to e.g., the presence of bright stars, globular clusters, and nearby galaxies) or incomplete coverage in the core ($griz$) DES imaging dataset \citep{sevilla21}. In regions with low galactic contamination, such systematic effects in optical surveys are not expected to be correlated with tSZ cluster detection. Therefore, to confirm additional tSZ candidates in regions masked by DES, we make use of the \citet[][hereafter WH24]{wen24} optical-infrared cluster catalog which adopts different masking choices than the DES analysis. We note, however, that a small number (4) of SPT candidate line of sight are so significantly contaminated by bright stars that confirmation is not possible with either catalog.

The WH24 cluster sample consists of 1.58 million clusters of galaxies identified via overdensities of stellar mass in spatial and photometric redshift space. 
The sample is produced by combining optical data from the 20,000 deg$^2$ Dark Energy Spectroscopic Instrument Legacy Surveys \citep{dey19} with infrared data from the  Wide-field Infrared Survey Explorer \citep[WISE;][]{wright10}. Redshift errors for clusters in  WH24 are quoted as $\sigma_z/(1+z)$ = 0.013 below $z=1$, with no uncertainties reported for higher-$z$ systems owing to limited spectroscopic calibrators.  

We search for WH24 counterparts within 1.5$\arcmin$ of tSZ candidate positions and, in the case of multiple matches, assign the richest counterpart to be the most probable association. The probability of spurious matches is computed as discussed in Section \ref{sec:contaminion}.  
We also leverage the WH24 coverage to flag candidates with significant secondary structures along the LOS. We identify secondary systems at redshifts at redshift differences $\Delta_z >3\sigma$ from our main associations and use the same threshold for spurious matches as for the primary assignments. 

Finally, we make use of spectroscopic redshifts from the literature where available. These redshifts are primarily sourced from targeted follow-up of previously identified SPT, ACT, and REFLEX clusters \citep[]{B_hringer_2001,Bleem_2015, Bayliss_2016, 2018A&A...620A...5A, 2019ApJ...870....7K}. We also use an archival redshift estimate from B24 for one high-redshift system, SPT$-$CLJ2344-6004, as it was not independently found in WH24 and is at too high a redshift $(z \sim 1.3)$ to be detected by redMaPPer.

\subsubsection{Contamination Fraction}\label{sec:contaminion}
We quantify the probability that an SPT detection candidate is a chance superposition of optical structures and CMB noise fluctuations as a function of cluster redshift and richness as the contamination fraction, $f_{\text{cont}}$:

\begin{equation}
f_{\text{cont}}(\lambda_i, z_i) = \frac{\int_{\lambda_i}^{\infty} f_{\text{rand}}(\lambda, z_i) \, d\lambda}{\int_{\lambda_i}^{\infty} f_{\text{obs}}(\lambda, z_i) \, d\lambda},
\end{equation}
where $f_{\text{rand}}$ and $f_{\text{obs}}$ are the richness distributions along random and candidate lines of sight \citep[]{2019MNRAS.488..739K, 10.1093/mnras/stae1359}. 
Estimation of this statistic using a cluster sample drawn from only 100 square degrees is noisy. 
To reduce the noise in these estimates we expanded the sample and area by making use of a tSZ candidate list constructed from the full 1500d SPT-3G footprint and analyzed in an identical fashion to this work.\footnote{This is at the expense of slightly overestimating the contamination as the full 1500d region does not benefit from the added depth of the SPTpol data.}
For the \spitzer-based confirmations, it is not possible to expand the sample in this way owing to the lack of \spitzer \ data over the full 1500d SPT-3G footprint. Instead, we follow B24 and adopt a more conservative confirmation criteria below.  

The contamination of the optically confirmed cluster sample is defined as:
\begin{equation}
    \text{contamination} = f_{\text{cont}}^{\text{max}} \times (1 - p(\xi > \xi_{\text{min}})),
    \label{eq:opticalcontamination}
\end{equation}
where $p(\xi > \xi_{\text{min}})$ is the purity of the tSZ candidate sample above $\xi_{\mathrm{min}}$. The purity and contamination of the SPT-Deep sample is calculated through simulated SPT-Deep-like maps described in Section \ref{sec:simulations}. As in B24, clusters are defined as confirmed for $f_{\text{cont}} < 0.2$ ($f_{\text{cont}} < 0.1$ for \spitzer \ confirmations).

\subsection{Cluster Mass Estimation}
\label{sec:cluster_mass_estimation}
The SPT detection significance-mass relation \citep{benson13} takes the form:

\begin{equation}
    \langle\text{ln}\ \zeta\rangle = \ln \left[ A_{\mathrm{SZ}}\left(\frac{M_{500}}{3 \times 10^{14} M_\odot h^{-1}}\right)^{B_{\mathrm{SZ}}} \left(\frac{E(z)}{E(0.6)}\right)^{C_{\mathrm{SZ}}} \right],
\label{eqn:mz}
\end{equation}
parameterized by the normalization $A_\mathrm{SZ}$, mass slope $B_\mathrm{SZ}$, and redshift evolution $C_\mathrm{SZ}$. A log-normal scatter, $D_\mathrm{SZ}$, on $\zeta$ is assumed, and $E(z) \equiv H(z)/H_0$. Here, $\zeta$ is the unbiased mass estimator that accounts for the maximization of $\xi$ over position and filter scales, defined as \citep{2010ApJ...722.1180V}:

\begin{equation}
    \zeta \equiv \sqrt{\left<\xi\right>^2 -3}\; \textrm{for} \left<\zeta\right>  > 2.
\end{equation}

The normalization of the $\zeta - M_{500}$ relation depends on the noise level of the field. As was done in previous SPT publications, $A_\mathrm{SZ}$ is rescaled as $\gamma \times A_\mathrm{SZ}$ for each individual SPT field to account for the changes in noise levels between SPT surveys. We calculate $\gamma$ by following  \citet{2024PhRvD.110h3509B} and fit the scaling parameters of Equation \ref{eqn:mz} to the SPT-Deep cluster abundances assuming a fixed $A_\mathrm{SZ}$ value following \citet{Bleem_2015}. For SPT-Deep we find a value of $\gamma = 4.97 \pm0.24$. We note that the original 150~GHz-only analysis on one SPT-SZ field \citep{Vanderlinde_2010} is defined to have a value of $\gamma = 1$, and for comparison, the SPTpol 500d survey has a value of $\gamma = 2.23$. This implies that clusters that are in both the SPT-Deep and SPTpol 500d catalogs will typically have a tSZ significance that is $\sim 2\times$  higher in SPT-Deep; similarly, clusters in common between SPT-Deep and SPT-SZ will typically have $\sim 5\times$ higher significance in SPT-Deep.

The mass for each galaxy cluster is then calculated through the posterior probability for mass
\begin{equation}
    P(M|\xi) = \frac{dN}{dM dz} \Bigg| _z P\left(\xi | M, z\right),
\end{equation}
where $\xi$ is the measured significance, $\frac{dN}{dM dz}$ is the assumed mass function \citep{2008ApJ...688..709T}, and $P\left(\xi | M, z\right)$ is the $\xi$-mass scaling relation defined above. Following previous SPT publications, the cosmological parameters for $\frac{dN}{dM dZ}$ are held fixed and only the scaling relation parameters are varied. The scaling relation parameters come from the \citet{Bleem_2015} catalog with updated redshifts presented in \citet{2019ApJ...878...55B} with best-fit parameters $A_\mathrm{SZ} = 4.08$, $B_\mathrm{SZ} = 1.69$, $C_\mathrm{SZ} = 0.87$, and $D_\mathrm{SZ} = 0.18$.

\subsection{Completeness}
\begin{figure}
  \includegraphics[width=0.45\textwidth]{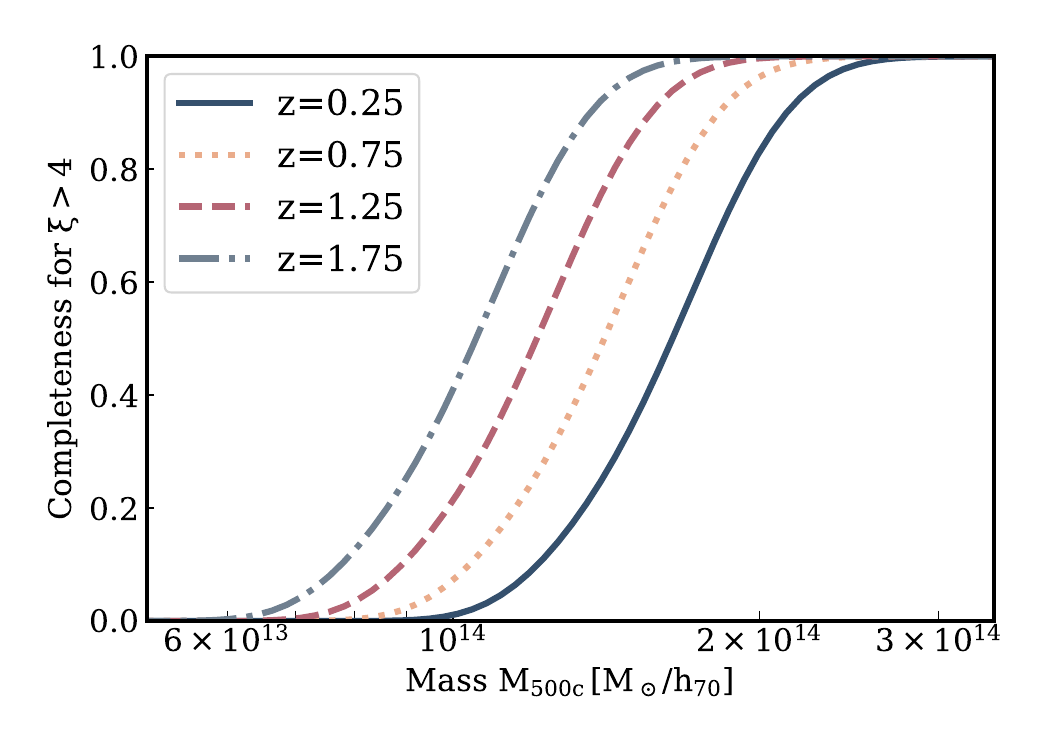}
  \caption{The expected completeness of the SPT-Deep cluster catalog as a function of mass at four different slices in redshift, $z =0.25, 0.75, 1.25, 1.75$. The sample is expected to be $> 90\%$ complete above \completenessmass ~at $z > 0.25$.}
  \label{fig:completeness}
\end{figure}

The completeness of the SPT-Deep catalog is calculated analytically and modeled as a Heaviside function in significance $\Theta(\xi - 4.0)$, reflecting the hard cut in tSZ significance used to select cluster candidates. This completeness in $\xi$ is converted to completeness in mass and redshift using the $\zeta - M$ relation discussed in Section \ref{sec:cluster_mass_estimation}, which now represents the probability of a cluster of a given mass at a given redshift to be found in the SPT-Deep cluster sample. This transformation includes both the intrinsic scatter from the $\xi - \zeta$ relationship and the observational scatter on $\xi$, modeled as a unit normal distribution. We find that the SPT-Deep catalog is expected to be $> 90\%$ complete at masses above \completenessmass~ at $z > 0.25$, shown in Figure \ref{fig:completeness}. We note that the completeness of the SPT-Deep catalog below a threshold of $z < 0.25$ becomes difficult to model owing to the filters applied during the SPT map-making process (discussed in depth in Section \ref{spt_map_making}) which remove large-scale Fourier modes, and hence large angular scale signals from clusters.

\section{SPT-Deep Simulated Maps and Sample Purity}
\label{sec:simulations}
To estimate the purity of the SPT-Deep cluster catalog, we construct 25 simulated maps with noise and source properties similar to the SPT-Deep maps. Construction of the simulated maps, as well as the catalog verification, follows previous SPT publications \citep[][B24]{Huang_2020}. We provide a brief overview, noting improvements in the simulation pipeline. 

\subsection{SPT-Deep Simulations}
To construct the SPT-Deep simulated maps, we begin by generating maps of the individual signal components. First, we construct Gaussian realizations of the following components:

\begin{itemize}
    \item CMB: Simulated CMB maps are constructed from CMB spectra from The Code for Anisotropies in the Microwave Background (CAMB; \citealp{2011ascl.soft02026L}) and the best-fit lensed \textit{Planck} 2018 \(\Lambda\)CDM primary CMB parameters \citep{Planck2020}.
    \item Dusty Sources: Dusty sources are modeled with two separate power spectra, a Poisson and a clustered term, using the best-fit spatial and spectral values from \citet{2021ApJ...908..199R}. At $\ell = 3000$ the amplitude of the Poisson term is measured to be $D_P^{3000} = 7.24 \pm 0.63 \mu K^2$ at 150~GHz; the one and two halo clustered terms are measured to be $D_{3000}^{1-\text{halo}} = 2.21 \pm 0.88 \mu K^2$ and  $D_{3000}^{2-\text{halo}} = 1.82 \pm 0.31 \mu K^2$. 
    \item kSZ: The kinetic-SZ effect is again modeled from the best fit \citet{2021ApJ...908..199R} power spectrum, which has an amplitude of $D_{kSZ}^{3000} = 3.0 \pm 1.0 \mu K^2$ at $\ell = 3000$ at 143~GHz. 
\end{itemize}

Identical to the astrophysical source treatment in the matched filter calculation, these spectra are rescaled to SPT-3G's effective frequencies. The rest of the simulated map components are not Gaussian realizations, which include:

\begin{itemize}
\item tSZ: Simulated clusters are added to the SPT-Deep maps through the use of the Outer Rim \citep{Heitmann_2019} Compton-$y$ maps. Covering a volume of $\sim 4\;\text{Gpc}^3$ with a mass resolution of $\sim 2.6 \times 10^{9} M_{\odot}$, Outer Rim is one of the largest N-body cosmology simulations to date. 25 independent, 100 deg$^2$ patches are selected from the full-sky Outer Rim simulation to ensure no overlap between simulated maps. These Compton-$y$ maps are converted into simulated maps of the tSZ at the SPT frequencies by rescaling the tSZ power spectrum to match the tSZ power measured in \citet{2021ApJ...908..199R} at $\ell = 3000$. 
\item Instrumental: The instrumental noise is calculated by creating signal-free coadds of individual map observations (by multiplying random halves of the observations by -1). These instrumental noise maps are masked with the same point source mask calculated for the minimum-variance cluster finder, as the intrinsic source variability of bright AGN can cause residuals in the calculated instrumental noise maps that show up as false positives in simulated data.
\end{itemize}

Finally, as described in Section \ref{sec:source_sub}, point sources may produce false cluster candidates. We note a significant change in the simulation pipeline with the handling of radio-galaxy point sources to validate our point-source cleaning process to maintain a high-purity cluster catalog.

\begin{itemize}
    \item Radio Galaxies: Simulated maps are populated with sources using the $\frac{dN}{dS}$ point source flux distribution from \citet{2011A&A...533A..57T}. Point source fluxes are measured at 150~GHz and scaled to the 95~GHz and 220~GHz maps with a spectral index of -0.7 and -0.9 respectively. The scatter on the spectral index is assumed to be 0.3 for both frequencies, following \citet{Everett_2020}. We populated sources up to a maximum flux cut of $50$ mJy at 150~GHz to match the observed interpolated flux cut in the SPT-3G maps. This is an improvement over B24 which did not populate maps with sources above the source subtraction threshold.
\end{itemize}

Cleaning of the simulated maps proceeds with the same methods detailed in Section \ref{sec:source_sub} to best mimic the SPT cluster extraction process. Simulated maps are processed with a matched filter optimized for point source detection to measure the significance of simulated sources at 95, 150, and 220~GHz as measured in SPT-3G maps. Sources with SNR $\geq 5$ at 95~GHz (equivalent to a flux cut of roughly 2~mJy) are subtracted from the maps, while sources above 6~mJy measured at 150~GHz are masked.

\begin{figure}
  \includegraphics[width=0.5\textwidth]{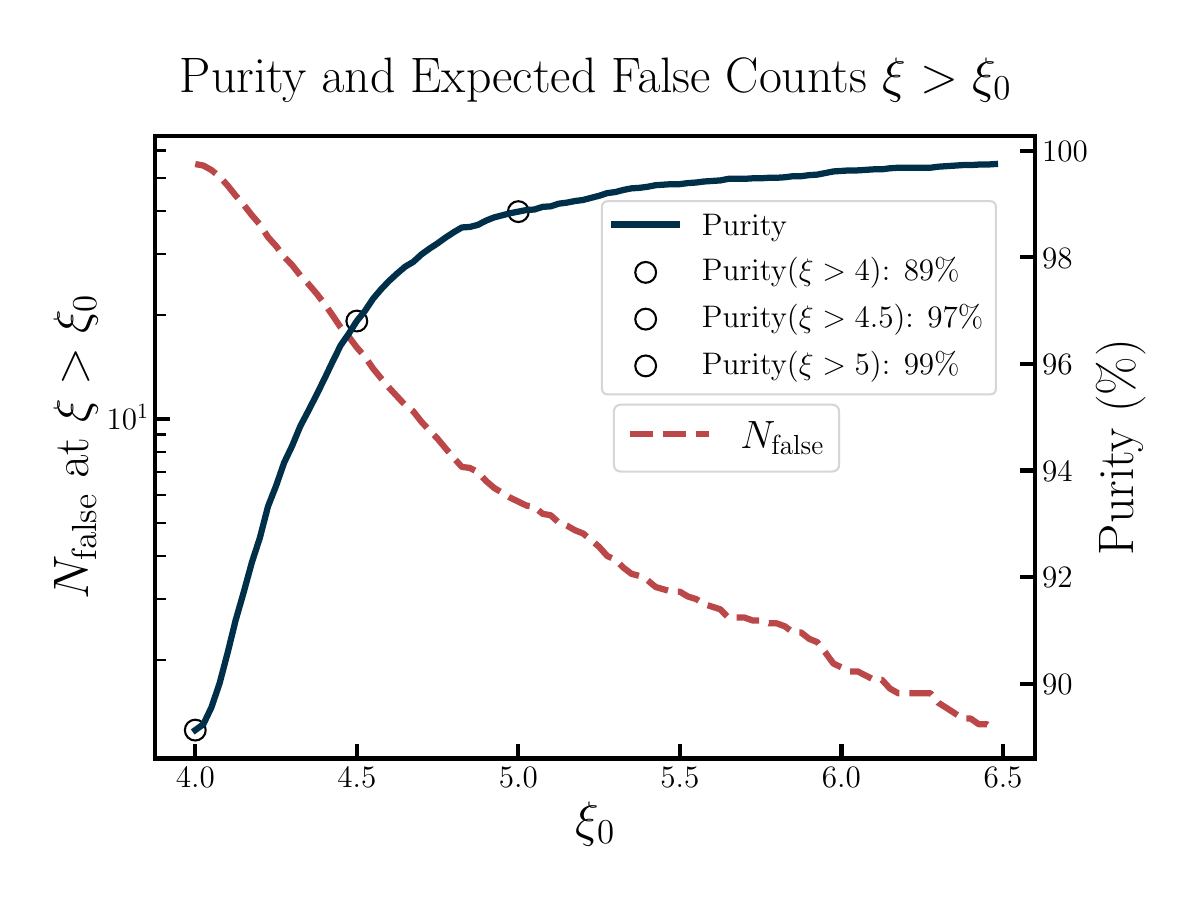}
  \caption{The number of false detections and purity of the SPT-Deep catalog based on a set of 25 simulated maps as a function of minimum signal-to-noise cut. We find a similar purity of $99\%$ above $\xi > 5$ as was found in previous SPT publications, and a purity of $89\%$ above the SPT-Deep minimum significance threshold of $\xi > 4$.}
  \label{fig:purity}
\end{figure}

\subsection{False Positive Estimation}
The number of false cluster detections as a function of candidate significance is calculated according to the method described in B24. We provide a brief overview of the process. 

The 25 simulated SPT-Deep maps are run through the cluster finder to produce a list of candidate detections. Candidates are matched to known simulated halo locations within a 2$'$ radius, beginning with the highest significance candidate. This process is repeated for 5000 random sight lines to determine the probability of random association with a halo of a given mass, $f_R(M)$. The number of false associations, $N_{\mathrm{FA}}$, above some minimum significance, $\xi_\textrm{min}$, is then calculated as:
\begin{equation}
    N_{\text{FA}}(M, \xi_{\text{min}}) = N_{\text{obs}}(M) - p(M, \xi_{\text{min}})N_{\text{cand}}.
\end{equation}
Where $N_\textrm{obs}(M)$ is the number of candidate associations, $N_\textrm{cand}$ is the number of cluster detections for each simulated map, and $p(M,\xi_{\text{min}})$ is the fraction of true
associations to total candidates, defined as:

\begin{equation}
p(M, \xi_{\text{min}}) = \frac{N_{\text{obs}}(M) - N_{\text{cand}}f_R(M)}{N_{\text{cand}}(1 - f_R(M))}.
\end{equation}

Owing to the steepness of the halo mass function, this quantity is computed in small bins in mass and summed to produce the total estimate of false associations. 

The purity of the final catalog is calculated by taking the difference between the total number of clusters in the catalog and $N_{\text{FA}}$, divided by the total number of clusters. The resulting purity of the catalog is shown in Figure \ref{fig:purity}. The results of our simulations show that the SPT-Deep catalog is $89\%$ pure above $\xi = 4$. This is consistent with the number of optically confirmed cluster candidates as described in Section \ref{sec:contaminion}. We note, however, that these results are sensitive to the accuracy of astrophysical source power estimation, as explored in B24. Combining the results of the simulated purity estimate with Equation \ref{eq:opticalcontamination}, we find that the purity of optically confirmed SPT-Deep cluster candidates is $\sim 98\%$ above $\xi > 4$ for a total of 433 true cluster detections. Given the simulated purity estimates a total of 445 real cluster candidates, this provides a rough estimate that 12 real cluster candidates are above the $f_{\text{cont}}$ threshold and failed to be optically confirmed.
\begin{figure*}
\centering
\includegraphics[width=\textwidth]{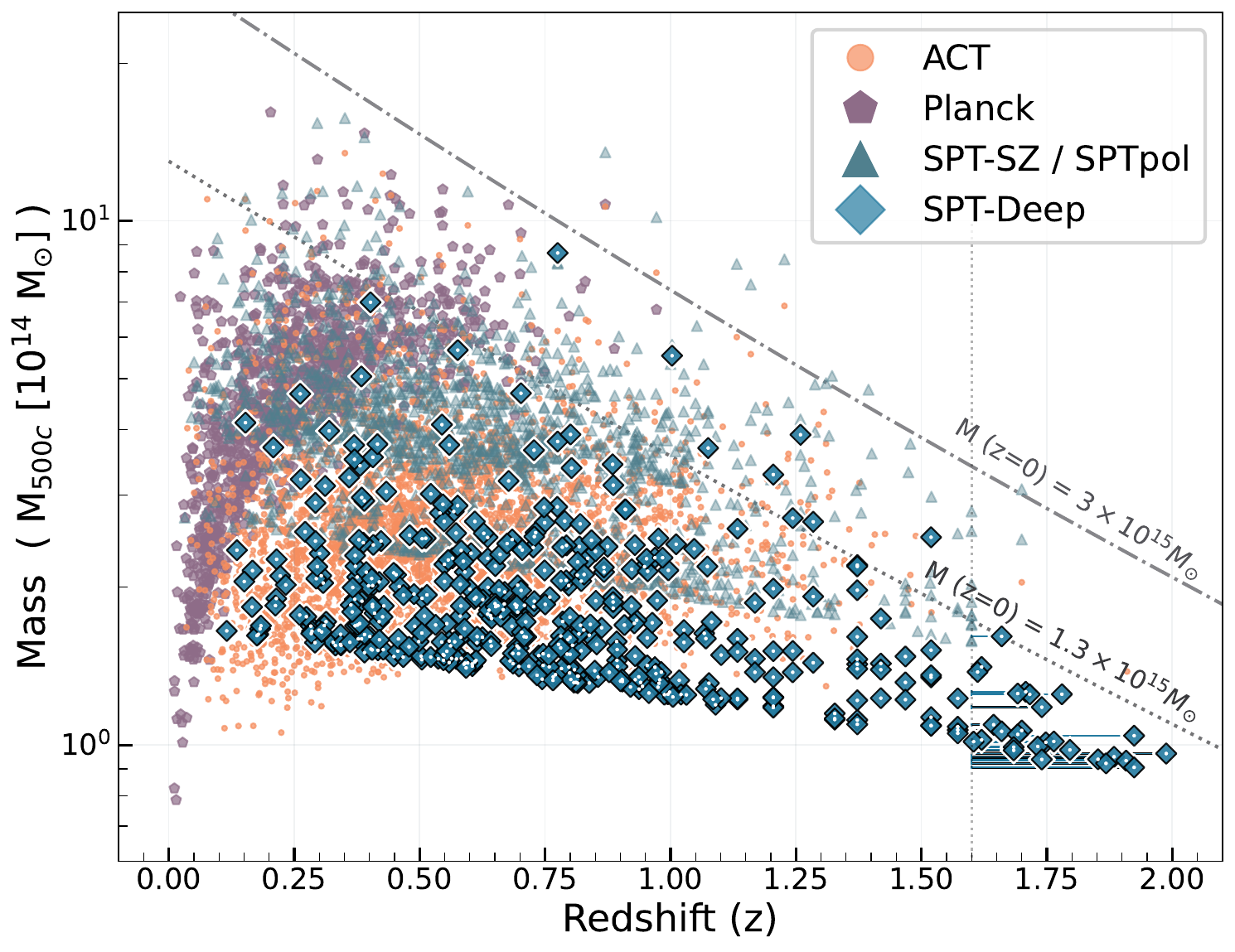}
  \caption{The SPT-Deep catalog on a $M_{500c}$ - $z$ plot compared to other SZ-selected cluster samples (ACT, Planck, SPT-SZ / SPTpol). We note there are \numclustersgtmax clusters ($\sim 6\%$ of the sample) with redshift $ z > 1.6$ that can only be reliably confirmed to be at least $z = 1.6$ as the [I1] - [I2] \spitzer ~color-color relation used to obtain IR redshifts flattens out above this redshift threshold. The redshifts shown for $z > 1.6$ clusters are drawn from a $N(z)$ halo-mass function probability distribution assuming a fixed cosmology. The upper dash-dotted line shows the mass-evolution of a $3 \times 10^{15} \msun$ cluster at z = 0 as a function of redshift for a fixed cosmology \citep{2010MNRAS.406.2267F}, while the lower dotted line represents the same mass-evolution for a $1.3 \times 10^{15} \msun$ cluster.} 
  \label{fig:mass_redshift_plot}
\end{figure*}

\section{The SPT-Deep Cluster Catalog}
\label{sec:min_var_catalog}
The SPT-Deep cluster catalog contains $\minvarnumcount$ galaxy cluster candidates, with $\crossmatchedclusters$ clusters confirmed with optical and/or IR data. In Table \ref{tab:data_column_keys} we describe the data released for each candidate in the SPT-Deep cluster catalog, which includes its location, redshift, mass, maximum detection significance $\xi$, and best-fit beta-model core size. The calculation of the mass and redshift estimates for cluster candidates is discussed in Section \ref{sec:external_datasets_characterization}. The SPT-Deep catalog has a median redshift of $z$ = $\medianredshift$,  spanning the redshift range $\minredshift < z < 1.6+$, and a median mass of $M_{500c} = \medianmass$ across $\minmass M_{\odot}/h_{70} < M_{500c} < \maxmass \times 10^{14} M_{\odot}/h_{70}$. We denote the upper limit of the SPT-Deep redshift range to be $z = 1.6+$ as $z = 1.6$ is the maximum redshift that the infrared color-redshift relation used to estimate cluster redshifts is reliable (see Section \ref{stellarbump}).

There are $\numclustersgteight$ confirmed clusters ($\percentclustersgeight\%$ of the sample) with $z > 0.8$,  $\numclustersgtone$ ($\percentclustersgtone\%$ of the sample) with $z > 1$, and \numclustersgtmax clusters ($\percentgtmax\%$ of the sample) with $ z > 1.6$. In total for the combined minimum-variance (Section \ref{sec:min_var_catalog}) and dust-nulled (see Section \ref{sec:cilc}) catalogs we confirm \confirmrm \ systems with redMaPPer, \confirmspitzer \ with \spitzer, and an additional \confirmarchive \ with WH24 and other archival sources. Comparing WH24 to our redMaPPer and \spitzer \ confirmed systems we find 278 of the redMaPPer confirmed clusters similarly confirmed in WH24 (i.e., with $f_{\text{cont}}<0.2$ and redshifts within 3$\sigma$ of the redMaPPer estimate) and 27 of the \spitzer \ systems also confirmed in that catalog. 

As the SPT-3G data is significantly deeper than both the SPTpol 100d and SPTpol 500d datasets, we also test the effect of an SPT-3G-only analysis against the full SPT-Deep sample. We find the addition of the SPTpol data increases the number of cluster candidates by $\sim 5\%$, with a median increase in measured significance of $3\%$.

The full mass and redshift distribution compared to similar SZ-selected catalogs is shown in Figure \ref{fig:mass_redshift_plot}. As discussed in e.g., \citet{Huang_2020}, the hard, minimum cut seen in mass in Figure \ref{fig:mass_redshift_plot} is a product of the hard cut in candidate significance at $\xi = 4$. The slope observed is produced by two effects: at low redshift, the angular size of clusters becomes comparable to large-scale CMB and atmospheric fluctuations which are aggressively filtered in the SPT-Deep maps. At higher redshifts the slope is primarily a product of the self-similar evolution of galaxy clusters. For display purposes, to more properly reflect the true redshift distribution of our sample, clusters with redshifts $z > 1.6$ are assigned a redshift by drawing from the expected $N(z)$ curve for a sample with our selection thresholds at our fiducial cosmology. These systems are reported at $z=1.6$ in the provided cluster sample table. 

\subsection{Comparisons to Other Catalogs}
We compare the results of the SPT-Deep catalog to similar cluster catalogs, particularly those from SPTpol 100d \citep{Huang_2020}, SPTpol 500d \citep{Bleem_2024}, ACT \citep[]{Hilton_2021}, and eROSITA \citep{2024A&A...685A.106B}. We cross-match the SPT-Deep catalog with external catalogs within a 2' radius and report the number of cross-matched clusters, as well as the median separation and median mass ratio of the cross-matched clusters.

\label{sec:catalog_comparioson}
\subsubsection{SPTpol}
To test consistency with previous results, we first compare the SPT-Deep catalog to two previously published catalogs from the SPTpol experiment. We find 120 clusters with matches in B24 in the SPT-Deep footprint with 115 cross-matched to SPT-Deep clusters. The matched candidates have a median separation of $0.34' \pm 0.02'$ and a median mass ratio of $1.07 \pm 0.02$. This does not include the statistical error on the field-scaling parameter $\gamma$ used to calibrate the $\zeta - M_{500}$ relation (See section \ref{sec:cluster_mass_estimation}). The remaining 5 clusters that are not cross-matched to SPT-Deep have a maximum significance of 4.2 and are likely false positives or systems whose tSZ signal scattered significantly high in the B24 dataset, which is consistent with purity estimates for the B24 catalog. In SPTpol 100d (for which cluster detections were reported at $\xi>4.6$, compared to $\xi>4.0$ in B24 and this work), we find 79 clusters lie in the SPT-Deep field with 78 cross-matches. The matched candidates have a median separation of $0.32' \pm 0.03$ and a median mass ratio of $1.02 \pm 0.02$. 

\subsubsection{ACT}
We cross-match the SPT-Deep catalog against the ACT DR-5 catalog and find 44 clusters that lie in the SPT-Deep footprint, with 34 cross-matches. Of the cross-matched clusters, we find a median separation of $0.49' \pm 0.05'$ and a median mass ratio of $1.07 \pm 0.04$. Of the 10 clusters not cross-matched, 1 lies in a masked region in SPT-Deep and the remaining clusters have a median significance of 4.3 in the ACT catalog with a maximum significance of 4.8. 

\subsubsection{eROSITA/eRASS1 Cosmology Catalog}
We find 22 clusters in the eROSITA/eRASS1 galaxy groups and clusters cosmology catalog \citep{2024A&A...685A.106B} that lie in the SPT-Deep footprint, with 18 cross-matches. Of the cross-matched clusters, we find a median separation of $0.49' \pm 0.09'$ and a median mass ratio of $0.96 \pm 0.05$. The cluster masses in the eROSITA catalog were estimated using a different cosmological model, particularly with a higher value of $\sigma_8$. Specifically, eROSITA assumed $\sigma_8 = 0.88$, whereas SPT-Deep adopts $\sigma_8 = 0.80$. We repeat our fit of $A_\mathrm{SZ}$ as described in Section \ref{sec:cluster_mass_estimation} for $\sigma_8 = 0.88$ and find the mass discrepancy increases to $0.89\pm0.05$. This mass discrepancy presents an interesting avenue for further detailed analysis, which we defer to future studies with the full SPT-3G 1500 deg$^2$ cluster catalog. Of the 4 eROSITA clusters not cross-matched, 3 lie in regions masked in SPT-Deep due to their proximity to radio sources. The one remaining cluster has a redshift of $z=0.11$ where filtering of SPT maps significantly decreases the completeness of the SPT-Deep catalog. 

\begin{figure*}
  \centering
  \includegraphics[width=5.75in]{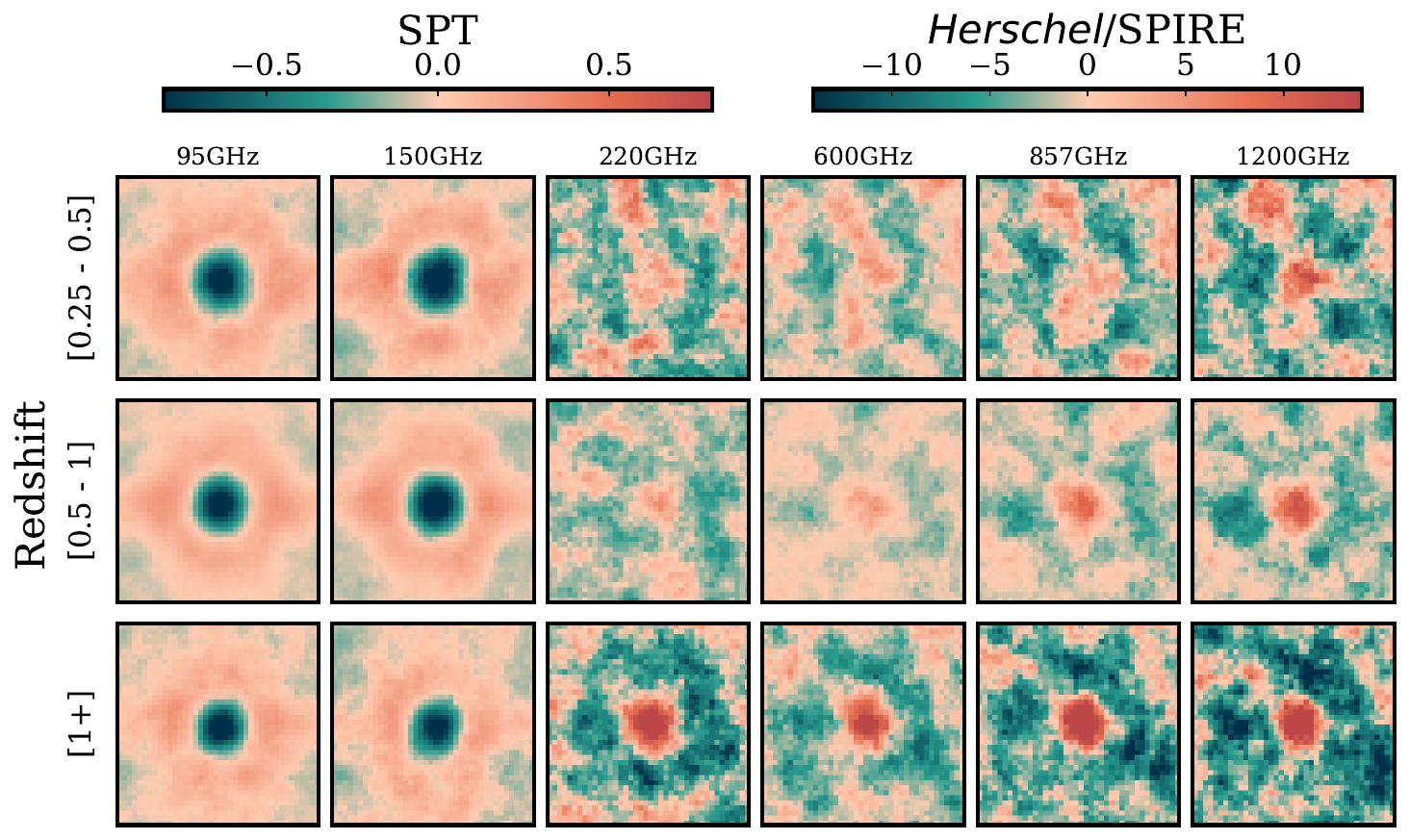}
  \caption{Median-weighted stacked cutouts of 422 confirmed clusters candidates (outlined in Section \ref{sec:contaminion}) with $z > 0.25$ in the SPT-Deep cluster catalog at 95, 150, 220, 600, 857, and 1200~GHz from the SPT-3G and \textit{Herschel}/SPIRE maps. We note a visual trends: the amplitude of the \fakecib increases with redshift, visible even in the tSZ null 220~GHz band. The amplitude of these stacks is normalized to the peak absolute value in the 95~GHz stack.}
  \label{fig:stacked_cutouts}
\end{figure*}

\section{Impact of Dust Contamination on Cluster Selection}
\label{sec:CIB}
The unprecedented depth of the SPT-Deep catalog enables the exploration of SZ-selected cluster populations at lower masses and higher redshifts than previously possible; we leverage these properties to investigate the impact of one potentially significant astrophysical contaminant in the high redshift regime---dust emission from cluster member galaxies---on the cluster selection function. We begin by exploiting the broad frequency coverage and depth of the SPT-Deep CMB maps, the sub-mm SPIRE maps, and the SPT-Deep cluster catalog to jointly fit a model of the tSZ and \fakecib at the location of clusters to estimate the magnitude of this contamination to the tSZ signal. The best-fit dust model is explicitly removed from the input CMB and SPIRE maps to produce an extended cluster catalog, optimizing the filtered map for cluster detection even in the presence of substantial dust contamination. In this section, we detail the dust contamination analysis, as well as the construction of the dust-nulled cluster catalog extension.

\subsection{Fitting for Dust Contamination of the tSZ signal}
To determine the spectral dependence of cluster-correlated dust emission, we first measure the flux in the SPT-Deep and \textit{Herschel}/SPIRE maps at the locations of confirmed cluster candidates in the SPT-Deep catalog. The steps to prepare the input maps are as follows: 

\begin{itemize}
    \item Input maps are converted to a common unit base of $\mathrm{Jy/sr}$; SPIRE maps are converted from their base units of $\mathrm{Jy/beam}$ by dividing them by the effective beam areas measured from \citet{Viero_2019}. SPT maps are converted from $\Delta T_{\mathrm{CMB, \mu K}}$ using the measured SPT-3G bandpasses.
    \item To minimize the impact of systematics such as cluster miscentering and to establish comparable flux measurements, beams are deconvolved from each individual map before they are reconvolved with the measured SPT-3G 150~GHz beam ($\sim 1.2'$). 
    \item Maps are filtered with the minimum-variance SPT-3G 150~GHz 0.25' matched-filter $\psi$ to optimize the signal-to-noise of high-redshift cluster-scale features. This mitigates the impact of contaminants such as large-scale CMB features and atmospheric/instrumental noise.
\end{itemize}

A sample of 422 confirmed clusters from the SPT-Deep catalog, located within the SPIRE footprint and with redshifts $z > 0.25$, is selected for the joint tSZ-dust fit, following the equation:

\begin{equation}
\begin{split}
    S_{\nu}^{\text{total}} = & \left[ y_{\text{tSZ}} \times I_0 \frac{x^4 e^x}{(e^x -1 )^2} f_{\textrm{SZ}}(x) \right] \\
    & + \left[ \frac{A_{\nu_0}^{\text{dust}} \left( \frac{\nu(1+z)}{\nu_0} \right)^{\beta_d} \times B_{\nu}[\nu(1 + z), T_{\text{d}}] }{B_{\nu_0}[T_{\text{d}}]} \right].
\end{split}\label{CIB_tSZ_fit_eq}
\end{equation}

The first bracketed term represents the tSZ contribution to the total flux, where $I_0 = \frac{2 (k_B T_{\mathrm{CMB}})^3}{(hc)^2}$. The second term represents the \fakecib with spectral index $\beta_d$, dust temperature $T_d$, Planck spectrum $B_{\nu}(\nu,T)$, and normalization amplitude $A_{\nu_0}^{\text{dust}}$, where $\nu_0$ is the normalization frequency of dust emission.

For each galaxy cluster, we construct a likelihood function to estimate three parameters: the temperature and amplitude of dust contamination, $T_d$ and $A_{\nu_0}^{\text{dust}}$, and the amplitude of tSZ signal, $y_{\text{tSZ}}$. We assume flat priors on all parameters and search the parameter range $T_{\mathrm{dust}} > 2.725$, $A_{\nu_0}^{\text{dust}} > 0$, and $y_{\text{tSZ}} > 0$. In our analysis, we fix the spectral index $\beta_d$ to a value of 2 \citep{Mak_2016}, as the degeneracy between the fitted dust parameters leads to poor constraints on the overall best-fit when floating $\beta_d$. However, we verify that $\beta_d = 2$ is an appropriate estimate of the spectral index by performing the tSZ-dust fit on clusters with redshifts $z > 1$ on a grid of $\beta_d$ values ranging from 1.2 to 2.6 in steps of 0.1, finding no significant change in the measured $\chi_{\mathrm{red}}^2$ of the resulting fits above $\beta_d \sim 1.8$.

To account for the evolution of the \fakecib with redshift, we divide our dataset into redshift bins: [0.25 $< z <$ 0.5], [0.5 $< z <$ 1], and [$z >$ 1], with 87, 233, and 102 clusters in each bin respectively. We note the removal of three clusters from this analysis: SPT-CL~J2353-5412 due to the presence of low-redshift galaxy contaminants, SPT-CL~J2332-5358 due to the presence of a lensed sub-mm galaxy, and the merging cluster SPT-CL~J2331-5052 (in which the close spatial proximity of two massive clusters biases the relative flux estimate, see e.g., Figure 6 in \citealt{Huang_2020}). In each bin, we simultaneously fit the likelihood functions of all clusters using a Markov chain Monte Carlo (MCMC) method. This approach allows us to jointly maximize the likelihood across all clusters in a given redshift bin, ensuring a more robust estimation of the fitting parameters. The best-fit parameters are obtained from the sum of the individual cluster log-likelihood functions. The errors for the measured flux at each frequency are estimated using bootstrapping, where points are randomly resampled with replacement from the input filtered maps, with the error calculated from the variation across the resampled datasets. We write the likelihood function as:

\begin{equation}
\ln \mathcal{L}(\boldsymbol{\theta}) = 
-\frac{1}{2} \sum_{i=1}^{N} \frac{\left[S_i - M(\boldsymbol{\theta}, \nu_i, z_i)\right]^2}{\sigma_i^2}.
\end{equation}

\begin{figure*}
  \centering
  \includegraphics[width=\textwidth]{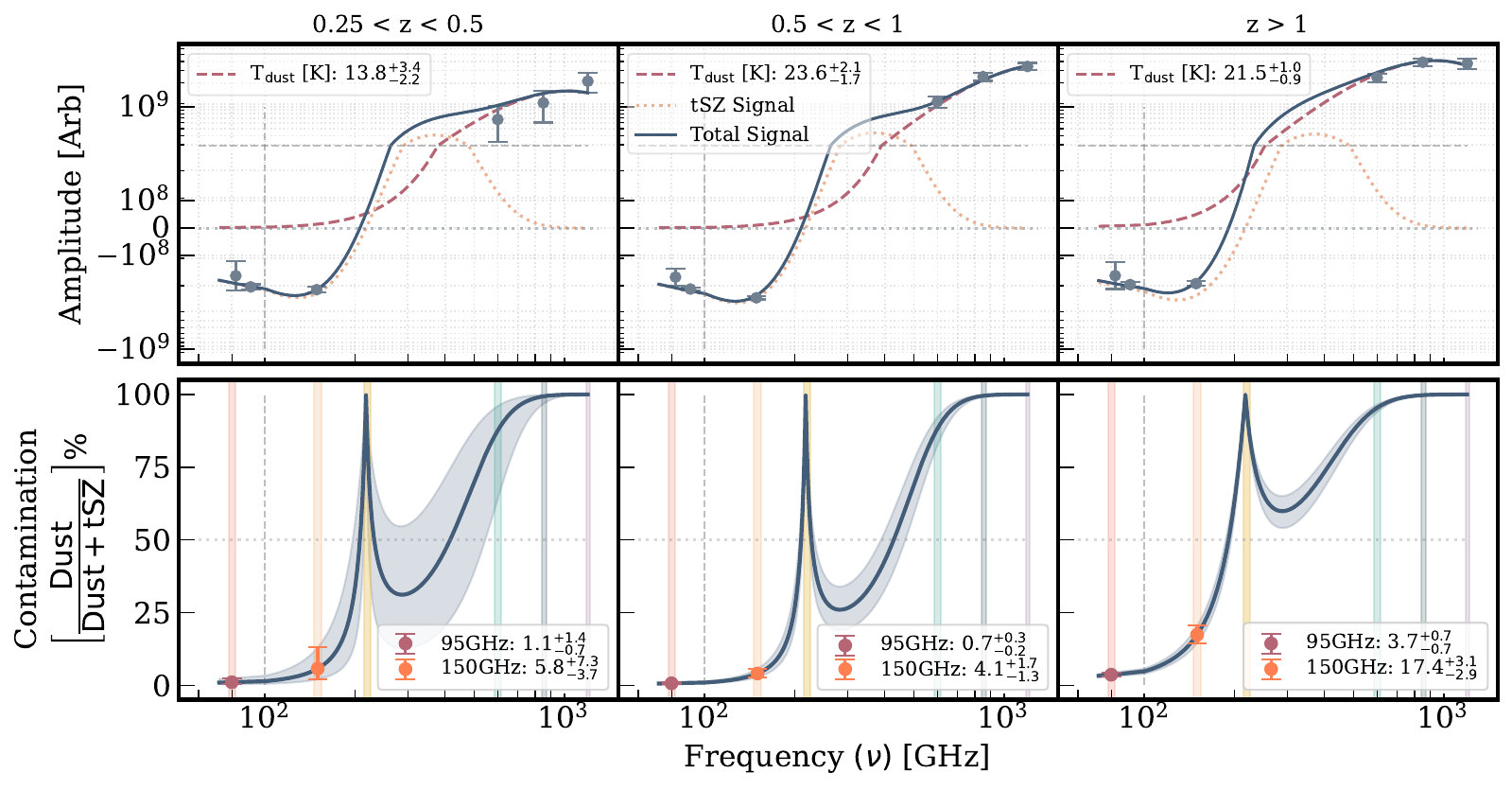}
  \caption{The results of the joint tSZ-dust fit on a subsample of 422 confirmed cluster candidates from the SPT-Deep catalog. We show the results in three redshift bins:  [0.25 $< z <$ 0.5], [0.5 $< z <$ 1], and [$z >$ 1], with 87, 233, and 102 clusters in each bin respectively. Plotted in the top panel is the mean signal at cluster locations in the three redshift bins at the 6 SPT and SPIRE frequencies with their bootstrapped error bars. The joint fit (blue) is shown alongside its dust contribution (red) and tSZ contribution (yellow). Both the x- and y- axes are log-linear for visualization purposes. In the bottom panel, we show the fractional dust contribution to the total measured flux as a function of frequency. Above the tSZ null at 220~GHz, where dust emission dominates the flux measurement, we observe a decline consistent with the expected contribution from the positive tSZ signal. Colored bars indicate the SPT and SPIRE central frequencies. For visualization purposes, the vertical and horizontal dashed lines in the upper and lower plots represent a transition between linear and log scaling.}
  \label{fig:cib_fit_plot}
\end{figure*}

Here, $S_i$ is the flux measured at cluster location $i$, $\boldsymbol{\theta} = \{T_d, A_{\nu_0}^{\text{dust}}, y_{\text{tSZ}}\}$, and we sum over the clusters in each redshift bin. The best-fit obtained from the MCMC analysis is illustrated in Figure \ref{fig:cib_fit_plot}. We find at $z < 1$, the dust signal is on-average $< \sim1\%$ and $<6\%$ of the tSZ signal at 95 and 150~GHz, respectively. As expected, we find a marked increase at $z > 1$ to $\sim 3\%$ and $\sim 17 \%$ at 95 and 150~GHz, respectively. We find best-fit temperatures for the dust to be \bestfittemplowz, \bestfittempmedz, and \bestfittemphighz ~K in the [0.25 $< z <$ 0.5], [0.5 $< z <$ 1] and [$z >$ 1] bins respectively, consistent with previous measurements of the temperature of the dust emission in clusters \citep[e.g.,][]{Soergel_2017,Erler_2018,2021MNRAS.502.4026F,Orlowski_Scherer_2021}. The best-fit parameters across all redshift bins can be found in Table \ref{table:CIB_results}. 

The results of our measurement of \fakecib at low redshifts are consistent with the negligible bias calculated by \citet{zubeldia2024planckszificataloguesnew} on a sample of low-redshift Planck clusters and from \citet{Orlowski_Scherer_2021} on a sample of ACT clusters, in agreement with observations of the low-redshift cluster environment being dominated by quiescent galaxies. Further studies on \fakecib  \citep{Bleem_2022, Bleem_2024} that span a larger range in redshift have provided weak evidence for a dust-induced bias on the tSZ signal, but lacked the depth to provide stringent constraints on the magnitude of this contamination. The low noise of the SPT-Deep maps, in conjunction with the large sample of high-redshift clusters from the SPT-Deep catalog, circumvents many limitations of previous analyses to provide the first constraining, empirically measured estimate of the \fakecib on tSZ selection as a function of redshift in clusters. This represents a large step in validating the selection function of future cosmological cluster catalogs to mitigate bias in resulting cosmological parameters.

We also test for any dependence of dust contamination of cluster mass by repeating our analysis with a new free parameter $\alpha$ in our fit, which scales the \fakecib with mass as:

{\small
\begin{equation}
\begin{split}
    S_{\nu}^{\text{total}} = & \left[ y_{\text{tSZ}} \left(\frac{M}{M_{\mathrm{median}}}\right)^{\frac{5}{3}} I_0 \frac{x^4 e^x}{(e^x -1 )^2} f_{\textrm{SZ}}(x) \right] +\\
    & \left[ \frac{A_{\nu_0}^{\text{dust}}  \left(\frac{M}{M_{\mathrm{median}}}\right)^{\alpha}  \left( \frac{\nu(1+z)}{\nu_0} \right)^{\beta} B_{\nu}[\nu(1 + z), T_{\text{d}}] }{B_{\nu_0}[T_{\text{d}}]} \right].
\end{split}
\end{equation}
}

We perform this fit on the 102 clusters in the $z > 1$ redshift bin, finding a value of $\alpha$ that is $1\sigma$ consistent with 0 at $\alpha = \highzslpha$. We repeat this fit, now applying a correction for Eddington bias to the flux measurements at 95~GHz and 150~GHz. The correction assumes that the flux bias scales with the mass bias, $M_{\text{bias}}$, as $M_{\text{bias}}^{5/3}$. In the limit where the flux at these frequencies is dominated by the tSZ signal, our results remain consistent with the nominal fit.

\begin{table}[ht]
    \centering
    \scriptsize 
    \begin{tabular}{ p{1.9cm} p{1.6cm} p{1.2cm} p{1.0cm} p{0.4cm}}
        \toprule
        \textbf{Redshift Bin} & \textbf{Temperature [K]} & \multicolumn{2}{c}{\textbf{Dust Contamination [\%]}} & $\chi^2_{\mathrm{red}}$\\
        \cmidrule(lr){3-4}
        & & \textbf{95~GHz} & \textbf{150~GHz} \\
        \midrule
        0.25 $< z <$ 0.50 & $13.8^{+3.4}_{-2.2}$  & $1.1^{+1.4}_{-0.7}$ & $5.8^{+7.3}_{-3.7}$ &1.40\\
        0.50 $< z <$ 1.0 & $23.6^{+2.1}_{-1.7}$   & $0.70^{+0.3}_{-0.2}$ & $4.1^{+1.7}_{-1.3}$& 2.11 \\
        $z >$ 1.0 & $21.5^{+1.0}_{-0.9}$  & $3.7^{+0.7}_{-0.7}$ & $17.4^{+3.1}_{-2.9}$ &1.72\\
        \bottomrule

    \end{tabular}
    \caption{The resulting best-fit dust temperature and fractional dust contamination in the [0.25 $< z <$ 0.5], [0.50 $< z <$ 1.0], and [$z  > 1.0$] redshift bins, alongside $\chi^2_{\mathrm{red}}$, the reduced chi-squared statistic calculated from the number of clusters in each bin and the three free fit-parameters. 
    }
    
    \label{table:CIB_results}
\end{table}

\subsection{The Constrained ILC}

\label{sec:cilc}
Emission from dusty galaxies reduces the significance of the tSZ signal used for finding cluster candidates. Traditional approaches to cluster finding (described in detail in Section \ref{sec:finding_clusters}) are optimized to reduce \fakecib across the entire map, but the spatial correlation of dust with the tSZ signal may make a global dust reduction insufficient to detect clusters in the presence of excess dust emission at cluster locations. An alternative approach is to filter maps using a constrained internal linear combination (cILC), which explicitly nulls an astrophysical source component of interest given its frequency dependence. Following  \citet{Bleem_2022}, the cILC filter can be written as:

\begin{equation}
    \psi_i (\nu_j, \ell) = \sum_{k,m} C_{\psi, ik} N^{-1}_{jm}(\ell)f_k(\nu_m)S_{\textrm{filt}}(\ell,\nu_m)
\end{equation}

where

\begin{equation}
    C^{-1}_{\psi,ij}(\ell) = \sum_{k,m} f_i(\nu_k)S_{\textrm{filt}}(\nu_k,\ell)N^{-1}_{km}(\ell)f_j(\nu_m)S_{\textrm{filt}}(\nu_m, \ell)
\end{equation}

Here, $f_i(\nu)$ is defined as the frequency dependence of signal $i$ at frequency $\nu$ and $C^{-1}_{\psi,ij}(\ell)$ captures the covariance between the null signal, dust, and the signal of interest, the tSZ. Dust contamination is modeled as a modified blackbody of the form given in Equation \ref{CIB_tSZ_fit_eq} with the best-fit parameters found in the ($z > 1$) redshift bin. 

The cILC cluster finding proceeds identically to the methods described for the minimum-variance case, save for the integration of the SPIRE maps into the matched filter {and a reduction in area due to the incomplete coverage of SPIRE over the SPT-Deep field. To incorporate these higher frequencies, we assume that the input SPIRE maps are primarily composed of dust emission. We therefore utilize their power spectrum to represent both the Poisson and clustered dusty galaxy noise terms, and assume all other astrophysical noise terms are negligible. The instrumental noise for the SPIRE maps is assumed to be white with magnitudes given by \citet{Viero_2019}. 

We find that the results of the cILC catalog are consistent with the minimum-variance catalog. The cILC catalog contains \numcilcclusters (\numcilcclustersgtfive) cluster candidates above $\xi > 4$ ($\xi > 5$) with \numcilcclustersconfirmed probabilistically confirmed candidates. Of the \ilccilccrossmatchnum candidates  cross-matched between the minimum-variance and cILC cluster catalog, we find a median significance ratio of $0.99$. This suggests that the noise penalty induced by nulling the measured dust spectrum is negligible with the addition of the SPIRE maps. For comparison, repeating this analysis with SPT data alone, we find a significance ratio of $0.96$, demonstrating that there is a small noise penalty of $4\%$ to measured cluster candidate significances using the SPT-only cILC maps. 

We find that the consistency between the cILC and minimum-variance cluster catalog is due to the power of the 220~GHz frequency band in subtracting dust emission. Sitting at the tSZ null, at small angular scales the 220~GHz frequency band is expected to be dominated by CIB/dust emission. As shown in Figure \ref{fig:psi_figured}, the 220~GHz frequency band subtracts off power across the $\ell$-ranges most sensitive to clusters. We compare the measured significance $\xi$ at the fiducial SPT-Deep cluster catalog locations in minimum-variance filtered SPT-3G, SPTpol~100d, and SPTpol~500d maps with and without the addition of the SPT-3G 220~GHz frequency band, finding a rough increase in significance of $\sim 20\%$ at $z > 1$ and fiducial $\xi > 4.5$ (compared to $\sim 14\%$ at $z < 1$), further validating the assumption that the 220~GHz frequency band is a powerful tracer of \fakecib for cluster selection. Our results suggest that, in regards to cluster finding, while infrared emission from cluster member galaxies does impact the cluster selection function at high redshift, the 220~GHz tSZ-null tracer is sufficient to mitigate its impact at the cluster mass ranges probed in this work. 

\section{Conclusion}
\label{sec:conclusion}
In this work, we present a new sample of galaxy clusters selected through their tSZ signal in the 100-square-degree SPT-Deep field. The mm-wave data in this work is drawn from three different surveys that contain the SPT-Deep field: the SPT-3G Main survey, the SPTpol 100d survey, and the SPTpol 500d survey. We describe improvements in the cluster selection pipeline, including an empirical quantification of point source contamination used to set source subtraction thresholds and improved cluster simulations for sample purity estimation techniques. 
The SPT-Deep catalog contains \minvarnumcount \  galaxy cluster candidates with $\xi > 4$ selected from coadded SPT-Deep maps with noise levels of 3.0, 2.2, and 9.0 $\mu \mathrm{K}_{\mathrm{CMB}-\mathrm{arcmin}}$ at 95, 150, and 220~GHz, respectively. Cluster candidates were confirmed by searching optical and infrared observations with the DES redMaPPer algorithm to identify significant red-sequence signatures and an IR-based code to search for IR galaxy over-densities at the candidate locations; we probabilistically confirm $\crossmatchedclusters$ cluster candidates. The median redshift of the SPT-Deep catalog is $z = \medianredshift$, and the median mass is $M_{500c} = \medianmass$, spanning a redshift range of $\minredshift < z \lesssim 1.8$ and a mass range of $\minmass  M_{\odot} / h_{70}< M_{500c} < \maxmass \times 10^{14}\ M_{\odot} / h_{70}$.

The sample is expected to be $> 90\%$ complete above \completenessmass ~at $z > 0.25$. A significant fraction of confirmed candidates lie at high redshift: \numclustersgteight clusters ($\percentclustersgeight\%$ of the sample) at $z > 0.8$, \numclustersgtone ($\percentclustersgtone\%$ of the sample) at $z > 1$, and \numclustersgtmax candidates at $z \geq 1.6$ ($\percentgtmax\%$ of the sample). The masses for confirmed cluster candidates are estimated from the $\xi - \mathrm{mass}$ relation which is calibrated by matching the abundance of observed clusters to a fixed $\Lambda \mathrm{CDM}$ cosmology. We find that for clusters in common between the SPT-Deep catalog and previous SPT cluster catalogs, the significance of the SPT-Deep catalog is roughly 2$\times$ larger than the SPTpol~500d catalog, or $5 \times$ larger than the SPT-SZ catalog. We produce SPT-Deep-like simulated maps to estimate the purity of the SPT-Deep catalog, which improves upon previous purity estimation techniques by directly incorporating the effects of source subtraction on cluster purity.

Given the high redshift range and low mass range sampled by the SPT-Deep catalog, we assess the impact of the \fakecib on the cluster selection function to understand the impact of astrophysical contamination on future cluster cosmology analyses. We perform a joint tSZ-dust fit on measurements of the central flux at the location of confirmed cluster candidates in the SPT-Deep field from maps spanning the frequency range 95 - 1200~GHz from SPT and \textit{Herschel}/SPIRE, fitting for the amplitude of the tSZ and the temperature and amplitude of dust. We find best-fit temperatures for the correlated IR emission in clusters to be approximately \bestfittemplowz, \bestfittempmedz, and \bestfittemphighz ~K in the [0.25 $< z <$ 0.5], [0.5 $< z <$ 1], and [$z >$ 1] bins respectively, broadly consistent with previous measurements. Most importantly, we find that at high redshift ($z > 1$), the tSZ is partially filled by $\cibconatminationhf$ ($\cibconatminationlf$) at 150~GHz (95~GHz) in the mass ranges probed by the SPT-Deep catalog. We also attempt to place constraints on the mass evolution of the dust amplitude at high redshift, finding results that are consistent with no evolution. 

We construct a dust-nulled map using the best-fit dust spectrum in the high-redshift ($z > 1$) bin and then repeat cluster detection to create an extended dust-nulled cluster catalog. We find consistent ($99\%$) estimates of the signal-to-noise of clusters between \ilccilccrossmatchnum cross-matched clusters. We attribute the consistency of the dust-nulled and minimum-variance cluster catalogs to the inclusion of the 220~GHz frequency band which is dominated by dust emission at small scales and is thus a powerful tracer of contaminating dust emission even in the minimum-variance catalog. The results of the empirical measurements of \fakecib on the tSZ signal at high redshifts highlight the necessity of a suitable tracer of dust emission in order to produce an unbiased sample of high-redshift clusters suitable for cosmology. 

The SPT-Deep catalog represents the first in a new generation of high-redshift, low-mass SZ cluster samples, marking a significant advancement for upcoming CMB experiments including the 1500 deg$^2$ and $10,000$ deg$^2$ SPT-3G \citep{osti_22167107, prabhu2024testingmathbflambdacdmcosmologicalmodel} surveys, the Simons Observatory \citep{Ade_2019}, and CMB-S4 \citep{abazajian2019cmbs4sciencecasereference}. These data sets will expand the number of high-redshift ($z>1$) massive clusters by over two orders of magnitude. 
Combined with unprecedented upcoming data from space-based infrared surveys such as the Euclid \citep{laureijs2011eucliddefinitionstudyreport} and Roman \citep{spergel2015widefieldinfrarredsurveytelescopeastrophysics} missions as well as ground-based data from Rubin Observatory's LSST \citep{2019ApJ...873..111I}, these samples will revolutionize cluster cosmology and astrophysical studies. 

\section*{Acknowledgments}

\facilities{Blanco (DECAM), NSF/US Department of Energy 10m South Pole Telescope (SPTpol, SPT-3G), \spitzer~ (IRAC)}, Herschel Space Observatory (SPIRE)

The South Pole Telescope program is supported by the National Science Foundation (NSF) through awards OPP-1852617 and 2332483. Partial support is also provided by the Kavli Institute of Cosmological Physics at the University of Chicago. Work at Argonne National Lab is supported by UChicago Argonne LLC, Operator of Argonne National Laboratory (Argonne). Argonne, a U.S. Department of Energy Office of Science Laboratory, is operated under contract no. DE-AC02-06CH11357.

This work is based in part on observations made with the \spitzer ~Space Telescope, which was operated by the Jet Propulsion Laboratory, California Institute of Technology under a contract with NASA.

Funding for the DES Projects has been provided by the U.S. Department of Energy, the U.S. National Science Foundation, the Ministry of Science and Education of Spain, 
the Science and Technology Facilities Council of the United Kingdom, the Higher Education Funding Council for England, the National Center for Supercomputing 
Applications at the University of Illinois Urbana-Champaign, the Kavli Institute of Cosmological Physics at the University of Chicago, 
the Center for Cosmology and Astro-Particle Physics at the Ohio State University,
the Mitchell Institute for Fundamental Physics and Astronomy at Texas A\&M University, Financiadora de Estudos e Projetos, 
Funda{\c c}{\~a}o Carlos Chagas Filho de Amparo {\`a} Pesquisa do Estado do Rio de Janeiro, Conselho Nacional de Desenvolvimento Cient{\'i}fico e Tecnol{\'o}gico and 
the Minist{\'e}rio da Ci{\^e}ncia, Tecnologia e Inova{\c c}{\~a}o, the Deutsche Forschungsgemeinschaft and the Collaborating Institutions in the Dark Energy Survey. 

The Collaborating Institutions are Argonne National Laboratory, the University of California at Santa Cruz, the University of Cambridge, Centro de Investigaciones Energ{\'e}ticas, 
Medioambientales y Tecnol{\'o}gicas-Madrid, the University of Chicago, University College London, the DES-Brazil Consortium, the University of Edinburgh, 
the Eidgen{\"o}ssische Technische Hochschule (ETH) Z{\"u}rich, 
Fermi National Accelerator Laboratory, the University of Illinois Urbana-Champaign, the Institut de Ci{\`e}ncies de l'Espai (IEEC/CSIC), 
the Institut de F{\'i}sica d'Altes Energies, Lawrence Berkeley National Laboratory, the Ludwig-Maximilians Universit{\"a}t M{\"u}nchen and the associated Excellence Cluster Universe, 
the University of Michigan, NSF NOIRLab, the University of Nottingham, The Ohio State University, the University of Pennsylvania, the University of Portsmouth, 
SLAC National Accelerator Laboratory, Stanford University, the University of Sussex, Texas A\&M University, and the OzDES Membership Consortium.

Based in part on observations at NSF Cerro Tololo Inter-American Observatory at NSF NOIRLab (NOIRLab Prop. ID 2012B-0001; PI: J. Frieman), which is managed by the Association of Universities for Research in Astronomy (AURA) under a cooperative agreement with the National Science Foundation.

The DES data management system is supported by the National Science Foundation under Grant Numbers AST-1138766 and AST-1536171.
The DES participants from Spanish institutions are partially supported by MICINN under grants PID2021-123012, PID2021-128989 PID2022-141079, SEV-2016-0588, CEX2020-001058-M and CEX2020-001007-S, some of which include ERDF funds from the European Union. IFAE is partially funded by the CERCA program of the Generalitat de Catalunya.

We  acknowledge support from the Brazilian Instituto Nacional de Ci\^encia
e Tecnologia (INCT) do e-Universo (CNPq grant 465376/2014-2).

This document was prepared by the DES Collaboration using the resources of the Fermi National Accelerator Laboratory (Fermilab), a U.S. Department of Energy, Office of Science, Office of High Energy Physics HEP User Facility. Fermilab is managed by Fermi Forward Discovery Group, LLC, acting under Contract No. 89243024CSC000002.

\bibliography{megaclusters}

\appendix

\section*{Affiliations}
\noindent
$^{1}${Department of Astronomy and Astrophysics, University of Chicago, 5640 South Ellis Avenue, Chicago, IL, 60637, USA}\\
 $^{2}${Kavli Institute for Cosmological Physics, University of Chicago, 5640 South Ellis Avenue, Chicago, IL, 60637, USA}\\
 $^{3}${High-Energy Physics Division, Argonne National Laboratory, 9700 South Cass Avenue., Lemont, IL, 60439, USA}\\
 $^{4}${Kavli Institute for Particle Astrophysics and Cosmology, Stanford University, 452 Lomita Mall, Stanford, CA, 94305, USA}\\
 $^{5}${Cerro Tololo Inter-American Observatory, NSF's National Optical-Infrared Astronomy Research Laboratory, Casilla 603, La Serena, Chile}\\
 $^{6}${School of Physics and Astronomy, Cardiff University, Cardiff CF24 3YB, United Kingdom}\\
 $^{7}${Laborat\'orio Interinstitucional de e-Astronomia - LIneA, Rua Gal. Jos\'e Cristino 77, Rio de Janeiro, RJ - 20921-400, Brazil}\\
 $^{8}${Department of Physics, University of Michigan, 450 Church Street, Ann Arbor, MI, 48109, USA}\\
 $^{9}${Fermi National Accelerator Laboratory, MS209, P.O. Box 500, Batavia, IL, 60510, USA}\\
 $^{10}${School of Physics, University of Melbourne, Parkville, VIC 3010, Australia}\\
 $^{11}${Center for Astrophysics  $|$ Harvard \& Smithsonian, Optical and Infrared Astronomy Division, Cambridge, MA 01238, USA}\\
 $^{12}${NIST Quantum Devices Group, 325 Broadway Mailcode 817.03, Boulder, CO, 80305, USA}\\
 $^{13}${Department of Physics, University of Colorado, Boulder, CO, 80309, USA}\\
 $^{14}${Institute of Cosmology and Gravitation, University of Portsmouth, Portsmouth, PO1 3FX, UK}\\
 $^{15}${Sorbonne Universit\'e, CNRS, UMR 7095, Institut d'Astrophysique de Paris, 98 bis bd Arago, 75014 Paris, France}\\
 $^{16}${University Observatory, Faculty of Physics, Ludwig-Maximilians-Universit\"at, Scheinerstr. 1, 81679 Munich, Germany}\\
 $^{17}${Department of Physics \& Astronomy, University College London, Gower Street, London, WC1E 6BT, UK}\\
 $^{18}${SLAC National Accelerator Laboratory, 2575 Sand Hill Road, Menlo Park, CA, 94025, USA}\\
 $^{19}${Kavli Institute for Astrophysics and Space Research, Massachusetts Institute of Technology, 77 Massachusetts Avenue, Cambridge, MA~02139, USA}\\
 $^{20}${Department of Physics, University of Chicago, 5640 South Ellis Avenue, Chicago, IL, 60637, USA}\\
 $^{21}${Enrico Fermi Institute, University of Chicago, 5640 South Ellis Avenue, Chicago, IL, 60637, USA}\\
 $^{22}${Instituto de Astrofisica de Canarias, E-38205 La Laguna, Tenerife, Spain}\\
 $^{23}${Universidad de La Laguna, Dpto. Astrofísica, E-38206 La Laguna, Tenerife, Spain}\\
 $^{24}${Institut de F\'{\i}sica d'Altes Energies (IFAE), The Barcelona Institute of Science and Technology, Campus UAB, 08193 Bellaterra (Barcelona) Spain}\\
 $^{25}${Department of Physics and McGill Space Institute, McGill University, 3600 Rue University, Montreal, Quebec H3A 2T8, Canada}\\
 $^{26}${School of Mathematics, Statistics \& Computer Science, University of KwaZulu-Natal, Durban, South Africa}\\
 $^{27}${University of Chicago, 5640 South Ellis Avenue, Chicago, IL, 60637, USA}\\
 $^{28}${Department of Physics, University of California, Berkeley, CA, 94720, USA}\\
 $^{29}${Jet Propulsion Laboratory, Pasadena, CA 91109, USA}\\
 $^{30}${Astronomy Unit, Department of Physics, University of Trieste, via Tiepolo 11, I-34131 Trieste, Italy}\\
 $^{31}${INAF-Osservatorio Astronomico di Trieste, via G. B. Tiepolo 11, I-34143 Trieste, Italy}\\
 $^{32}${Institute for Fundamental Physics of the Universe, Via Beirut 2, 34014 Trieste, Italy}\\
 $^{33}${Dunlap Institute for Astronomy \& Astrophysics, University of Toronto, 50 St. George Street, Toronto, ON, M5S 3H4, Canada}\\
$^{34}${David A. Dunlap Department of Astronomy \& Astrophysics, University of Toronto, 50 St. George Street, Toronto, ON, M5S 3H4, Canada}\\
 $^{35}${Universit\'e Paris-Saclay, Universit\'e Paris Cit\'e, CEA, CNRS, AIM, 91191, Gif-sur-Yvette, France}\\
 $^{36}${Department of Astronomy, University of Illinois Urbana-Champaign, 1002 West Green Street, Urbana, IL, 61801, USA}\\
 $^{37}${Physics Division, Lawrence Berkeley National Laboratory, Berkeley, CA, 94720, USA}\\
 $^{38}${Centro de Investigaciones Energ\'eticas, Medioambientales y Tecnol\'ogicas (CIEMAT), Madrid, Spain}\\
 $^{39}${Department of Physics, IIT Hyderabad, Kandi, Telangana 502285, India}\\
 $^{40}${Canadian Institute for Advanced Research, CIFAR Program in Gravity and the Extreme Universe, Toronto, ON, M5G 1Z8, Canada}\\
 $^{41}${Joseph Henry Laboratories of Physics, Jadwin Hall, Princeton University, Princeton, NJ 08544, USA}\\
 $^{42}${Department of Astrophysical and Planetary Sciences, University of Colorado, Boulder, CO, 80309, USA}\\
 $^{43}${California Institute of Technology, 1200 East California Boulevard., Pasadena, CA, 91125, USA}\\
 $^{44}${Department of Physics, University of Illinois Urbana-Champaign, 1110 West Green Street, Urbana, IL, 61801, USA}\\
 $^{45}${Department of Physics and Astronomy, University of California, Los Angeles, CA, 90095, USA}\\
 $^{46}${Department of Physics and Astronomy, Michigan State University, East Lansing, MI 48824, USA}\\
 $^{47}${Institute of Theoretical Astrophysics, University of Oslo. P.O. Box 1029 Blindern, NO-0315 Oslo, Norway}\\
 $^{48}${Faculty of Physics and Astronomy, University of Missouri--Kansas City, 5110 Rockhill Road, Kansas City, MO 64110, USA}\\
 $^{49}${Center for AstroPhysical Surveys, National Center for Supercomputing Applications, Urbana, IL, 61801, USA}\\
 $^{50}${Harvey Mudd College, 301 Platt Boulevard., Claremont, CA, 91711, USA}\\
 $^{51}${Instituto de Fisica Teorica UAM/CSIC, Universidad Autonoma de Madrid, 28049 Madrid, Spain}\\
 $^{52}${Department of Physics, University of Cincinnati, Cincinnati, OH 45221, USA}\\
 $^{53}${Department of Physics and Astronomy, University of Pennsylvania, Philadelphia, PA 19104, USA}\\
 $^{54}${Department of Physics \& Astronomy, University of California, One Shields Avenue, Davis, CA 95616, USA}\\
 $^{55}${European Southern Observatory, Karl-Schwarzschild-Str. 2, 85748 Garching bei M\"{u}nchen, Germany}\\
 $^{56}${Department of Physics, Stanford University, 382 Via Pueblo Mall, Stanford, CA, 94305, USA}\\
 $^{57}${Universit\"at Innsbruck, Institut f\"ur Astro- und Teilchenphysik, Technikerstr. 25/8, 6020 Innsbruck, Austria}\\
 $^{58}${Department of Physics and Astronomy, Northwestern University, 633 Clark St, Evanston, IL, 60208, USA}\\
 $^{59}${Institute for Computational Cosmology, Durham University, South Road, Durham DH1 3LE, UK}\\
 $^{60}${Centre for Extragalactic Astronomy, Durham University, South Road, Durham DH1 3LE, UK}\\
 $^{61}${School of Mathematics and Physics, University of Queensland,  Brisbane, QLD 4072, Australia}\\
 $^{62}${Santa Cruz Institute for Particle Physics, Santa Cruz, CA 95064, USA}\\
 $^{63}${Center for Cosmology and Astro-Particle Physics, The Ohio State University, Columbus, OH 43210, USA}\\
 $^{64}${Department of Physics, The Ohio State University, Columbus, OH 43210, USA}\\
 $^{65}${LPSC Grenoble - 53, Avenue des Martyrs 38026 Grenoble, France}\\
 $^{66}${Harvard-Smithsonian Center for Astrophysics, 60 Garden Street, Cambridge, MA, 02138, USA}\\
 $^{67}${Max-Planck-Institut f\"{u}r extraterrestrische Physik,Giessenbachstr.\ 85748 Garching, Germany}\\
 $^{68}${Department of Physics, University of California, One Shields Avenue, Davis, CA, 95616, USA}\\
 $^{69}${Department of Physics, Case Western Reserve University, Cleveland, OH, 44106, USA}\\
 $^{70}${Australian Astronomical Optics, Macquarie University, North Ryde, NSW 2113, Australia}\\
 $^{71}${Lowell Observatory, 1400 Mars Hill Rd, Flagstaff, AZ 86001, USA}\\
 $^{72}${Departamento de F'isica Matem'atica, Instituto de F'isica, Universidade de S\~ao Paulo, CP 66318, S\~ao Paulo, SP, 05314-970, Brazil}\\
 $^{73}${Laborat'orio Interinstitucional de e-Astronomia - LIneA, Rua Gal. Jos'e Cristino 77, Rio de Janeiro, RJ - 20921-400, Brazil}\\
 $^{74}${George P. and Cynthia Woods Mitchell Institute for Fundamental Physics and Astronomy, and Department of Physics and Astronomy, Texas A\&M University, College Station, TX 77843,  USA}\\
 $^{75}${Instituci\'o Catalana de Recerca i Estudis Avan\c{c}ats, E-08010 Barcelona, Spain}\\
 $^{76}${Excellence Cluster Universe, Boltzmannstr.\ 2, 85748 Garching, Germany}\\
 $^{77}${Department of Astrophysical Sciences, Princeton University, Peyton Hall, Princeton, NJ 08544, USA}\\
 $^{78}${Materials Sciences Division, Argonne National Laboratory, 9700 South Cass Avenue, Lemont, IL, 60439, USA}\\
 $^{79}${Observat\'orio Nacional, Rua Gal. Jos\'e Cristino 77, Rio de Janeiro, RJ - 20921-400, Brazil}\\
 $^{80}${Hamburger Sternwarte, Universit\"{a}t Hamburg, Gojenbergsweg 112, 21029 Hamburg, Germany}\\
 $^{81}${School of Physics and Astronomy, University of Minnesota, 116 Church Street SE Minneapolis, MN, 55455, USA}\\
 $^{82}${Laboratoire de physique des 2 infinis Ir\`ene Joliot-Curie, CNRS Universit\'e Paris-Saclay, B\^at. 100, Facult\'e des sciences, F-91405 Orsay Cedex, France}\\
 $^{83}${Department of Physics and Astronomy, Pevensey Building, University of Sussex, Brighton, BN1 9QH, UK}\\
 $^{84}${Department of Physics, Yale University, P.O. Box 208120, New Haven, CT 06520-8120}\\
 $^{85}${Department of Physics, Northeastern University, Boston, MA 02115, USA}\\
 $^{86}${Kavli Institute for Astrophysics and Space Research, Massachusetts Institute of Technology, 70 Vassar St, Cambridge, MA 02139}\\
 $^{87}${Liberal Arts Department, School of the Art Institute of Chicago, 112 South Michigan Avenue, Chicago, IL,60603, USA }\\
 $^{88}${Argelander-Institut f\"ur Astronomie, Auf dem H\"ugel 71, D-53121 Bonn, Germany}\\
 $^{89}${Three-Speed Logic, Inc., Victoria, B.C., V8S 3Z5, Canada}\\
 $^{90}${School of Physics and Astronomy, University of Southampton,  Southampton, SO17 1BJ, UK}\\
 $^{91}${Department of Physics, Faculty of Science, Chulalongkorn University, 254 Phayathai Road, Pathumwan, Bangkok Thailand. 10330}\\
 $^{92}${Vera C. Rubin Observatory Project Office, 933 North Cherry Avenue, Tucson, AZ 85721, USA}\\
 $^{93}${Computer Science and Mathematics Division, Oak Ridge National Laboratory, Oak Ridge, TN 37831}\\
 $^{94}${Space Science and Engineering Division, Southwest Research Institute, San Antonio, TX 78238}\\
 $^{95}${Department of Astronomy, University of California, Berkeley,  501 Campbell Hall, Berkeley, CA 94720, USA}\\
 $^{96}${STAR Institute, Quartier Agora - All\'ee du six Ao\^ut, 19c B-4000 Li\`ege, Belgium}\\
 $^{97}${Center for Astrophysics | Harvard \& Smithsonian, 60 Garden St, Cambridge, MA 02138, USA}\\

\clearpage
\begin{table}[ht]
\centering
\label{tab:data_column_keys}
\begin{tabular}{@{}p{7cm}p{3cm}p{8cm}@{}}
\toprule
\textbf{Column} & \textbf{Units} & \textbf{Description} \\
\midrule
\midrule
\toprule
SPT\_ID &  & The SPT Cluster Candidate Name \\
\midrule
RA & degrees (J2000) & Right Ascension returned by tSZ cluster finder \\
\midrule
DEC & degrees (J2000) & Declination returned by tSZ cluster finder  \\
\midrule
XI\_ILC / XI\_cILC &  & Detection significance $\xi$ in the minimum-variance cluster catalog (ILC) or dust-nulled catalog (cILC), if applicable \\
\midrule
THETA\_CORE\_ILC / THETA\_CORE\_cILC & arcminutes (') & Matched Filter Scale $\theta_c$ corresponding to $\xi_{\textrm{ILC}}$ or $\xi_\textrm{cILC}$, if applicable \\
\midrule
REDSHIFT &  & Redshift of optical/IR galaxy over density counterpart \\
\midrule
REDSHIFT\_SOURCE &  & Source of cluster candidate redshift estimate\\
\midrule
 REDSHIFT\_UNC &  & Redshift uncertainty \\
\midrule
SPECZ &  & Flagged 1 if spectroscopic redshift  \\
\midrule
M500 & $10^{14} M_{\odot}/h$ & Halo mass defined where the mean density is 500 times the critical density of the universe. \\
\midrule
M500\_UERR & $10^{14} M_{\odot}/h$ & 1 sigma upper uncertainty on mass \\
\midrule
M500\_LERR & $10^{14} M_{\odot}/h$ & 1 sigma lower uncertainty on mass \\
\midrule
LAMBDA &  & Richness $\lambda$ of optical/IR galaxy over-density \\
\midrule
CONTAMINATION &  & Integrated optical/IR contamination $> \lambda$\\
\midrule
CONFIRMED &  & Flagged 1 if contamination is $<$ contamination confirmation threshold \\
\midrule
LOS &  & If there is a secondary structure along the line of sight (LOS) with $f_\textrm{cont}<0.2$\\
\midrule
LOS\_Z &  & Redshift secondary structure\\
\midrule
LOS\_LAMBDA &  & Richness $\lambda_{\textrm{LOS}}$ secondary structure\\
\midrule
LOS\_FCONT &  & Integrated optical/IR contamination $> \lambda_{\textrm{LOS}}$\\
\midrule

COMMENT &  & Additional information about the cluster candidate (if applicable)\\

\bottomrule
\end{tabular}
\centering
\caption{\makebox[\textwidth][l]{Description of the Columns in the Primary Cluster Catalog FITS File}}

\end{table}

\end{document}